# Mathematical modeling and control of a tilt-rotor aircraft

Xinhua Wang[1] and Lilong Cai[2]

[1] Department of Mechanical and Aerospace Engineering,
Monash University, Melbourne, VIC 3800, Australia Email: wangxinhua04@gmail.com

[2] Department of Mechanical and Aerospace Engineering,
Hong Kong University of Science and Technology, Hong Kong, China

**Abstract:** This paper presents a novel model of large-size tilt-rotor aircraft, which can operate as a helicopter as well as being capable of transition to fixed-wing flight. Aerodynamics of the dynamic large-size tilt-rotors based on blade element method is analyzed during mode transition. For the large-size aircraft with turboshaft engines, the blade pitch angles of the rotors are regulated to vary according to the desired level of thrust, and the following expressions are formulated explicitly: rotor thrust and blade pitch angle, drag torque and blade pitch angle. A finite-time convergent observer based on Lyapunov function is developed to reconstruct the unknown variables and uncertainties during mode transitions. The merits of this design include the modeling of dynamic large-size tilt-rotor, ease of the uncertainties estimation during the tilting and the widely applications. Moreover, a switched logic controller based on the finite-time convergent observer is proposed to drive the aircraft to implement the mode transition with invariant flying height.

## 1 INTRODUCTION

Helicopters and fixed-wing airplanes have their advantages and shortcomings. Helicopters can take off and land vertically, but they cannot fly forward in high speed, and their payloads are very limited comparing to fixed wing plane with the same gross weight [1-4]. On the other hand, conventional fixed-wing aircrafts can fly forward in high speed and large payloads. However, they cannot take off and land vertically, and the appropriate runways are required.

There are several ways to perform the vertical takeoff and landing (VTOL) maneuver such as tilting rotor, tilting wing, thrust vectoring, tail sitter, tilting fuselage, flapping wing, multi-propeller multifunction, etc. There are some types of thrust vectoring aircrafts, such as manned aircrafts AV-8B Harrier [5] and F-35 [6]. These aircrafts are designed for actual combat. The reason why these aircrafts can perform vertical takeoff and landing is due to their jet engines with tilt jettubes. Although they are powerful, the jet gas is very hot and harmful, and they can easily destroy the ground environment or inflict injuries to people nearby. These aircrafts are not suitable for many civil and rescue operations. Moreover, such VTOL aircraft with jet engines is less efficient in hover than a conventional helicopter or a tilt-rotor aircraft of the same gross weight [7].

One of the classical tilt-rotor aircrafts is V-22 [8]. Similarly the Bell Eagle Eye UAV [9] which is also based on tilt-rotor technology has a large success in the civil and military domains. However, this design brings its own problems, since the degradation in stability is usually observed in high speed forward flight (airplane mode) [10]. Moreover, the involved equations of motion are highly coupled and nonlinear. The fundamental characteristics of V-22, which includes a very high roll moment of inertia, result in a tendency to pilot-induced oscillation. Extreme attention must be paid to preventing over control in the roll channel. Moreover, the roll channel is easily affected by pitch dynamics. V-22 is a large "U"-shaped structure with two very large masses at the ends (the engines). The result is the need for long and flexible runs of hydraulic, electrical, and mechanical lines that are more susceptible to tensile loads, bending loads, abrasions, and deformations than conventional helicopter designs. Because of the high disk-loading (129.63 kg/m²), the downwash velocity of V-22 is about twice that of any conventional helicopter. And because of the side-by-side placement of the prop-rotors, there are two distinct downwash wakes that are transverse to the flight direction. This has several operational implications that bear on safety issues.

Tail-sitters, as the name implies, sit on their tail when not in flight [11-14]. They take off and land vertically, making them a member of the VTOL family of aircraft. Equipped with a powerful engine, tail-sitters can utilize a "prop-hanging" technique to hover in place. Additionally, they can make transition to a level flight mode and fly in a conventional fixed-wing mode, which is more energy efficient than the hover mode. These aircrafts have higher rotor disk loadings. A novel unmanned tail-sitter aircraft called SkyTote has been designed [15-18]. It was originally conceived as an airborne conveyor belt that would use a VTOL capability to minimize ground handling. The concept demonstrator is a 'tail-sitter' configuration and utilizes coaxial, counter-rotating rotors. A relatively large cruciform tail provides directional control in the airplane modes as well as serving as landing gear in the helicopter mode. However, the flying height varies when the aircraft undergoes a vertical to horizontal transition. The aircraft climbs upward during mode transition. Importantly, a large thrust force should be provided to finish the mode transition. Moreover, the aerodynamics of the complex mechanical drive device, i.e., coaxial counter-rotating propellers, is difficult to be analyzed during mode transition.

A tilt-fuselage aircraft was presented in [19] to keep the flying height invariant during mode transition. It is a rotor-fixed wing aircraft with two free wings. During mode transition, the fuselage is tilted, and the free wings are kept at a given angle of attack. However, it is difficult to analyze the aerodynamics of the tilt fuselage during mode transition.

In [20], a quad tilt-wing unmanned aerial vehicles (UAV) were presented, which is capable of vertical takeoff and landing (VTOL) like a helicopter and long duration horizontal flight like an airplane. The aircraft is electric powered and small size. During mode transition, all the wings and the quad rotors are tilted. In [21], a multi-rotorcraft with morphing capabilities was presented. The aircraft is able to rotate its body into a vertical flying mode configuration. The middle base, in which the inertial sensors are fixed, is always horizontal during the morphing stage. However, the servo motor is difficult to keep the middle base horizontal during mode transition. Accordingly, the attitude measurement will be affected adversely. The two aircrafts in [20, 21] are all electric powered and small size.

The electric powers are usually suitable to drive small-size aircrafts. For an electric powered tilt-rotor aircraft, the rotor speeds are allowed to freely vary according to the desired level of thrust. However, for a large-size aircrafts with turboshaft engines, the rotor speeds freely varying according to the desired level of thrust is unrealistic for current state-of-the art technology for conventional rotorcrafts, i.e., the control of the thrust output is impossible to be achievable by changing the rotor speed. The aerodynamics of large-size tilt-rotor is more complex than that of small-size tilt-rotor during mode transition. Moreover, the tilt-rotor aircrafts are underactuated mechanical systems, which exhibit high nonlinear and time-varying behaviors. Meanwhile, the influences of aerodynamic disturbance, unmodelled dynamics and parametric uncertainties are not avoidable in modeling. These nonlinearities and uncertainties render great challenges in the design of flight control system.

Nonlinear modeling and analysis for mechanical systems have been presented in recent years [22]. To overcome these problems above, in this paper, a novel model of large-size tilt-rotor aircraft is presented, which can operate as a helicopter as well as being capable of transition to fixed-wing flight. The control of the thrust outputs is achieved by changing the blade pitch angles of the rotors, and the rotor speeds are fixed in rated power. The aerodynamics of the large-size tilt-rotor based on blade element method is analyzed during mode transition, and the following relations are formulated explicitly: rotor thrust and blade pitch angle, rotor drag torque and blade pitch angle. The designed tilt-rotor aircraft is more suitable to large-size aircraft applications. Its structure is based on quad-rotor aircrafts design [23-25]. Quad-rotor aircraft is a system consisting of four individual rotors attached to a rigid cross frame. It is an omnidirectionally symmetrical VTOL vehicle ideally suited to stationary and quasi-stationary flight conditions. Therefore, when the tilt-rotor aircraft hovers, takes off or lands, the control methods of quad-rotor aircraft can be used directly. During the mode transition from hover to forward flight and vice versa, the quad rotors are tilted synchronously by a gearing driven arrangement. Moreover, a finite-time convergent observer is developed to reconstruct the unknown variables and uncertainties during mode transitions. Based on the observer, a switched logic controller is proposed to drive the aircraft to implement the mode transition with the invariant flying height. Comparing with the electric powered tilt-rotor aircrafts, the modeling of the presented tilt-rotor aircraft is large size, the uncertainties during mode transition are estimated easily, and it has the widely applications with much long-time cruise and larger payload.

## 2 TILT-ROTOR AIRCRAFT DESIGN

### 2.1 Configuration of aircraft

The designed tilt-rotor aircraft is shown in Figs. 1(a)-1(d).

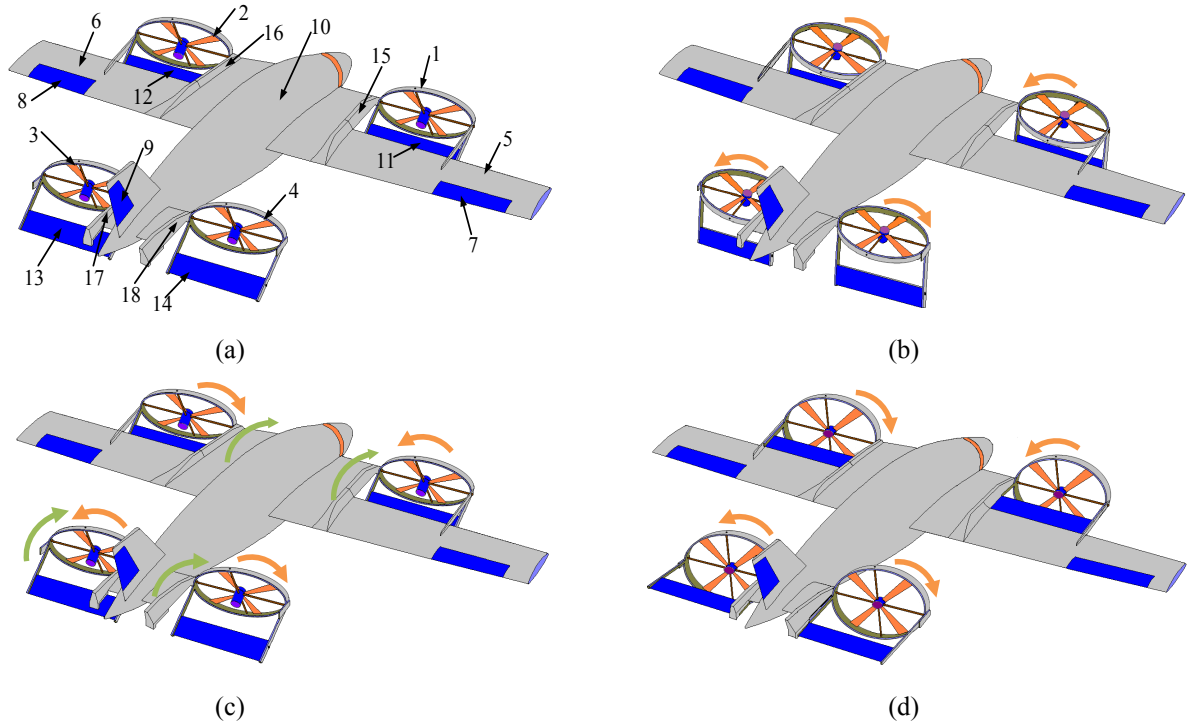

Fig. 1. Tilt-rotor aircraft: (a) Aircraft structure, (b) Modes of hover, takeoff and landing, (c) Mode transition from hover to forward flight and vice versa, (d) Mode of forward flight

**Mechanical components of tilt-rotor aircraft in Fig. 1(a):**

1: front-right rotor, 2: front-left rotor, 3: rear-left rotor, 4: rear-right rotor, 5: right fixed wing, 6: left fixed wing, 7: right aileron, 8: left aileron, 9: vertical tail, 10: fuselage, 11: front-right free wing, 12: front-left free wing, 13: rear-left free wing, 14: rear-right free wing, 15: front-right gearing arrangement, 16: front-left gearing arrangement, 17: rear-left gearing arrangement, 18: rear-right gearing arrangement

For the tilt-rotor aircraft, there are the following four flight modes:

*Hover:* When the aircraft is climbing, it is put on to at a certain height, normally out ground effect where the thrusts of the quad rotors compensate the helicopter weight *mg* (See Figure 1 (b)). The attitude is regulated by changing the blade pitch angles of the quad rotors, and the rotor speeds are fixed in rated power. Thus the control of the thrust and torque outputs is achievable. A control method for quad-rotor aircraft can be used directly.

*Vertical fight:* This fight mode starts when the aircraft is at rest on the ground in ground effect. Then takeoff is produced and the aircraft climbs. Vertical descent precedes landing. In the absence of perturbations the thrusts of the quad rotors are always vertical (See Figure 1 (b)).

*Mode transition:* It is considered that this fight mode is out of ground effect. At the beginning, the aircraft is assumed to be in hover (see Figure 1(b)). As the gearing arrangement is driven, the quad rotors are tilted synchronously and pitch towards the horizontal, which in turn causes the horizontal speed of the aircraft to increase (see Figure 1(c)). The attitude is controlled by changing the blade angles of the quad rotors to make the fixed wings obtain a given angle of attack in accordance with the relative wind. The gravity of the aircraft is counteracted mainly by the vertical force of the thrust generated by the rotors, and the height of the aircraft is kept invariant. With the horizontal speed increasing, the fixed wings develop lift. The aircraft is made transition into horizontal flight in a fixed wing mode (see Figure 1(d)). During this mode transition, before the tilt angle attains a given value, roll dynamics is controlled by the thrusts differential of the left and right rotors. Yaw dynamics is regulated by the counteractive moments brought by the rotation of the rotors in the air. Pitch dynamics is controlled by the thrusts differential of the front and rear rotors. After the tilt angle exceeds a given value, the flap bias angle of rear two free wings is selected to control pitch dynamics. Roll dynamics are controlled by the counteractive moments brought by the rotation of the rotors in the air. Yaw dynamics are regulated by the thrusts differential of the left and right rotors. For level-to-hover transition, the reverse procedure is used. As the gearing arrangement is driven, the quad rotors are tilted backward. It causes the horizontal speed of the aircraft to decrease and the vertical thrust vector gradually increases to keep the height invariant. Thus, the aircraft slows and performs transition to hover.

*Forward fight:* We also consider that this fight mode will be out ground effect. The thrust of the quad rotors ensures forward fight, and the lift force generated by fixed wings keeps the height of the aircraft invariant (See Figure 1 (d)).

**2.2 Gearing arrangement for tilt rotors**

A synchronous gear-driven arrangement for the tilt-rotor aircraft is designed to implement mode transition, as shown in Figure 2(a). The directions of the drive arrows denote the hover-to-level transition. In the gear-driven arrangement, the symmetrical structure shown in Figure 2(b) is adopted. With this gear-driven arrangement, the quad rotors can be tilted synchronously from vertical to level and vice versa.

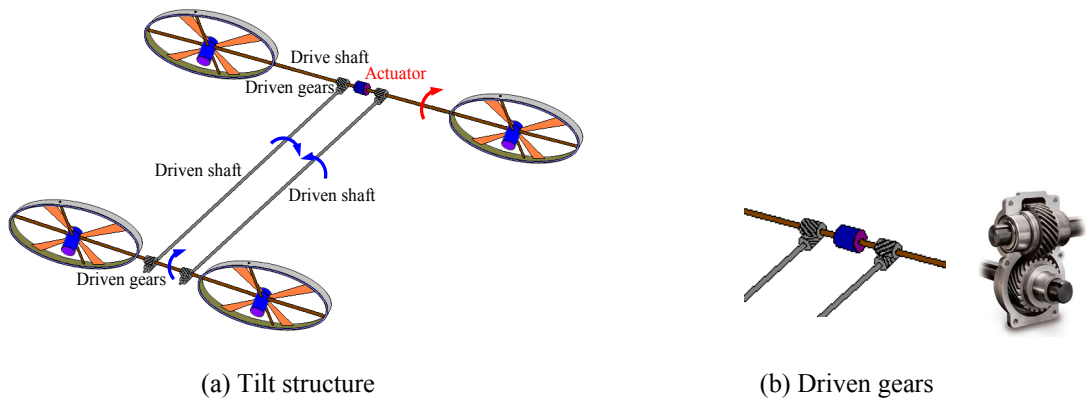

(a) Tilt structure          (b) Driven gears

Fig. 2. Tilt structure and driven gears

**2.3 Gross weight of aircraft**

In this paper, let the gross weight of aircraft be $m$, and four same turboshaft engines are selected, the weight of each engine is $m_e$, and its thrust-to-weight ratio is $\rho_e = T_e/(m_e g)$, where $g$ is the gravity acceleration, $T_e$ is rotor thrust in rated power. In order to consider the control maneuverability $\rho_{mT}$ during hover or mode transition, the thrust $4T_e$ should be larger than the gross weight $mg$ of aircraft. We select $\rho_{mT} = 4T_e/mg = \sqrt{2}$. Therefore, each rotor thrust can be carried out as $T_e = \rho_{mT} mg / 4$.

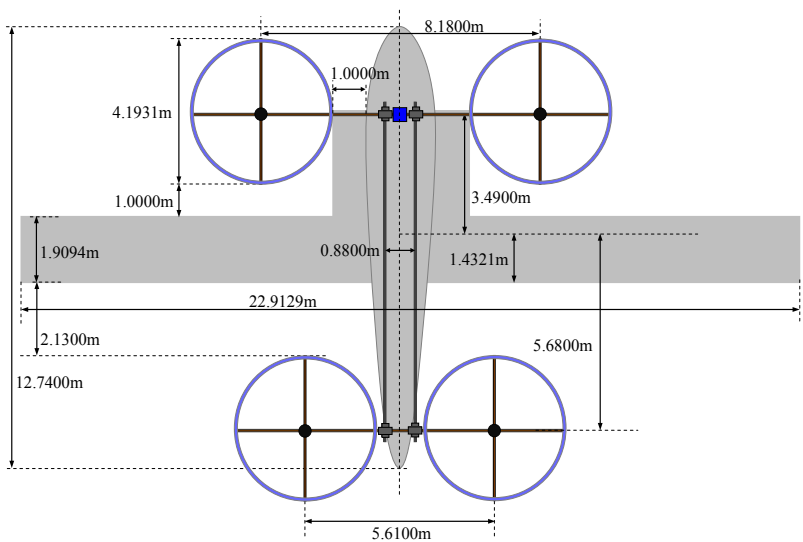

Fig. 3. Size of tilt-rotor aircraft

### 2.4 Rotor radius

Assume the thrust force generated by the $i$-th rotor is $T_i$ ($i = 1, 2, 3, 4$), and its thrust coefficient $C_{Ti}$ and rotor power coefficient $C_{Pi}$ are defined, respectively, as [7]

$$C_{T_i} = 2T_i / \left(\rho \Omega_i^2 R^2 A\right) \tag{1}$$

$$C_{P_i} = 2P_i / \left(\rho \Omega_i^3 R^3 A\right) \tag{2}$$

where, $A$ is the disk area, $\Omega_i$ is the rotating speed of the $i$-th rotor, $R$ is rotor radius, $P_i$ is the rotor power, and $\rho$ is the air density. It is known that $P_i = T_i v_i$, where $v_i$ is the induced velocity of rotor. Therefore, power coefficient $C_{Pi}$ can be written as

$$C_{P_i} = \frac{2T_i v_i}{\rho A (\Omega_i R)^3} = \frac{2T_i}{\rho A (\Omega_i R)^2} \frac{v_i}{\Omega_i R} = C_{T_i} \lambda_i \tag{3}$$

where $\lambda_i = v_i / (\Omega_i R)$ is the induced inflow ratio in hover. From (1), we obtain

$$\lambda_i = \frac{v_i}{\Omega_i R} = \frac{1}{\Omega_i R} \sqrt{\frac{T_i}{2\rho A}} = \sqrt{\frac{2T_i}{4\rho A (\Omega_i R)^2}} = \frac{1}{2} \sqrt{C_{T_i}} \tag{4}$$

Therefore, from (3) and (4), the ideal power can be written as

$$C_{P_i} = C_{T_i} \frac{1}{2} \sqrt{C_{T_i}} = \frac{1}{2} C_{T_i}^{3/2} \tag{5}$$

In hovering flight, the induced power predicted by the simple momentum theory can be approximately described by a modification to the momentum result in (5), i.e.,

$$C_{P_{i1}} = \frac{1}{2} \kappa C_{T_i}^{3/2} \tag{6}$$

where $\kappa$ is called an induced power correction factor. The profile power of rotor can be obtained as follow:

$$P_0 = \Omega_i p \int_0^R D y dy \tag{7}$$

where $p$ is the number of blades and $D$ is the drag force per unit span at a section on the blade at a distance $y$ from the rotational axis. The profile power consumed by the rotor requires the drag coefficients of the airfoils that make up the rotor blades. The drag force can be expressed conventionally as

$$D = 0.5\rho U^2 b C_d = 0.5\rho(\Omega_i y)^2 b C_d \tag{8}$$

where $b$ is the blade chord, and $U$ is the resultant velocity at the disk. The section profile drag coefficient $C_d$ is assumed to be constant, i.e., $C_d = C_{d0}$. Then the profile power is deduced to

$$P_0 = 0.5\rho p \Omega_i^3 b C_{d_0} \int_0^R y^3 dy = \frac{1}{8}\rho p \Omega_i^3 b C_{d_0} R^4 \tag{9}$$

Therefore, the profile power coefficient $C_{p0}$ can be written as

$$C_{P_0} = \frac{2P_0}{\rho A(\Omega_i R)^3} = \frac{\frac{1}{4}\rho p \Omega_i^3 b C_{d_0} R^4}{\rho \pi R^2 (\Omega_i R)^3} = \frac{1}{4}(\frac{pb}{\pi R})C_{d_0} = \frac{1}{4}(\frac{pbR}{\pi R^2})C_{d_0} = \frac{1}{4}\sigma C_{d_0} \tag{10}$$

where rotor solidity is expressed as $\sigma = (pbR)/(\pi R^2) = (pb)/(\pi R)$, and it has been defined as the ratio of total blade area to the disk area. Armed with the estimates of the induced power coefficient (6) and the profile power coefficient (10), the rotor power by using the modified momentum theory results that

$$C_{p_i} = C_{P_{i1}} + C_{P_0} = \frac{1}{2}\kappa C_{T_i}^{3/2} + \frac{1}{4}\sigma C_{d_0} \tag{11}$$

From Eqs. (1), (2) and (11), the ratio $P_i/T_i$ is given by

$$\frac{P_i}{T_i} = \Omega_i R \frac{C_{P_i}}{C_{T_i}} = \Omega_i R (\frac{1}{2}\kappa C_{T_i}^{1/2} + \frac{1}{4}\frac{\sigma C_{d_0}}{C_{T_i}}) \tag{12}$$

By differentiating Eq. (12) with respect to $C_{Ti}$, it is shown that for a rotor with rectangular blades the operating $C_{Ti}$ to give the lowest $T/P$ is

$$0 = \Omega_i R (\frac{1}{4}\kappa C_{T_i}^{-1/2} - \frac{1}{4}\frac{\sigma C_{d_0}}{C_{T_i}^2}) \tag{13}$$

i.e.,

$$C_{T_i} = (\sigma C_{d_0}/\kappa)^{2/3} \tag{14}$$

From (1) and (14), the disk loading for maximum power loading is at

$$DL = \frac{T_i}{A} = \frac{mg}{4\pi R^2} = 0.5 C_{T_i}\rho(\Omega_i R)^2 = 0.5\rho(\Omega_i R)^2 \left(\frac{\sigma C_{d0}}{\kappa}\right)^{2/3} \tag{15}$$

It is assumed that for design purposes each rotor of the quad rotors carries a quarter of the total weight of the aircraft. Profile drag coefficient $C_{d0}$ is assumed be constant and independent of Reynolds number $R_e$ and Mach number $M$. This equation determines the optimum radius of the rotor to maximize power loading at a given gross weight. Solving for the rotor radius in Eq. (15) gives

$$R = \sqrt{mg/(\pi DL)}/2 \tag{16}$$

Similarly, for two-rotor aircrafts, such as V-22, we obtain the rotor radius as follow: $R_{2r} = \sqrt{mg/(2\pi DL)}$. Under the same disk loading requirement, we obtain $R/R_{2r} = 1/\sqrt{2}$.

With respect to the conventional two-rotor aircrafts, the rotor radius of the presented tilt-rotor aircraft is shorter. For different VTOL rotorcrafts, they have different disk loadings: CH-47D Chinook (1 rotor, 2 blades), 14 kg/m²; Bell 206B3

JetRanger (1 rotor, 2 blades), 18 kg/m²; CH-47D Chinook (2 rotors, 3 blades), 43 kg/m²; Mil Mi-26 (1 rotor, 7 blades), 71 kg/m²; CH-53E Super Stallion (1 rotor, 7 blades), 72 kg/m²; MV22B Osprey, 129.63 kg/m². (2 rotors, 3 blades). A suitable disk loading $DL=DL_{QT} = 60$kg/m² is required for the presented quad tilt-rotor aircraft. Therefore, from Eq. (16), the rotor radius of the tilt-rotor aircraft can be computed as follow:

$$R = \sqrt{mg/(\pi DL_{QT})}/2 \tag{17}$$

## 2.5 Number of blades and Rotor solidity

The selection of the number of blades (for a given blade area or solidity) is usually based on dynamic rather than aerodynamic criteria, that is, it is based on the minimization of vibratory loads, which is easier for rotors that use a larger number of blades. An analysis of the data [26] suggests that hover performance is primarily affected by rotor solidity σ; the number $p$ of blades is secondary. Based on the requirement of the blade aspect ratio, we can select the blade aspect ratio $\rho_b$. Then from Eq. (17), we obtain the blade chord $b$. Values of $\sigma$ for contemporary helicopters vary from about 0.06 to 0.12. Lager helicopters generally tend to have upper solidity rotors. Here, we expect the rotor solidity is $\sigma=(pb)/(\pi R)$. Therefore, the number of blades is

$$p = \sigma\pi R/b \tag{18}$$

## 2.6 Fixed-wing loading

A large wing aspect ratio $A_w=12$ is selected for the fixed wing in order to obtain a long-time cruise. Moreover, we expect the desired wing loading $WL_w$ after comparing with other fixed wing aircrafts. Therefore, the wing area is obtained:

$$S_w = m/WL_w \tag{19}$$

Accordingly, the wing span $l_w$ and chord $c_w$ for the fixed wing are obtained, respectively:

$$l_w = \sqrt{A_w S_w}, c_w = \sqrt{S_w/A_w} \tag{20}$$

In addition, considering of the effect of the free wings during cruise, we expect the desired wing loading $WL_{ws}$. Then the sum of wings area is calculated as:

$$S = S_w + S_{front} + S_{fi} \tag{21}$$

where $S_w$, $S_{front}$ and $S_{ft}$ are the areas of fixed wing, front wing and free wings, respectively. Therefore, the area of each free wing is: $S_{fi}=(m/WL_{ws}-S_w-S_{front})/4$. Other sizes for the aircraft are shown in Figure 3.

## 3 MATHEMATICAL MODEL

In this section, by analyzing the aerodynamics of mode transition, a 6-degree of freedom (DOF) nonlinear model of the tilt-rotor aircraft is established. Moreover, the uncertainties in the system dynamics are considered.

### 3.1 Mathematical model during mode transition

The forces and torques of the aircraft during mode transition are shown in Fig. 4.

#### 3.1.1 Coordinates and frames

In Fig. 4, $C$ is the centre of gravity of the aircraft. $M_\beta$ is the pitch moment generated by gearing arrangement. Let

i) $\Gamma_g=(E_x, E_y, E_z)$ denote the right handed inertial frame;

ii) $\Gamma_b=(E_x^b, E_y^b, E_z^b)$ denote the frame attached to the aircraft's fuselage whose origin is located at its center of gravity;

iii) $\Gamma_\beta=(E_x^\beta, E_y^\beta, E_z^\beta)$ denote the right handed inertial frame of the tilt-rotor aircraft;

iv) $\Gamma_r=(X_r, Y_r, Z_r)$ denote the right handed inertial frame of blade element (See Fig. 5).

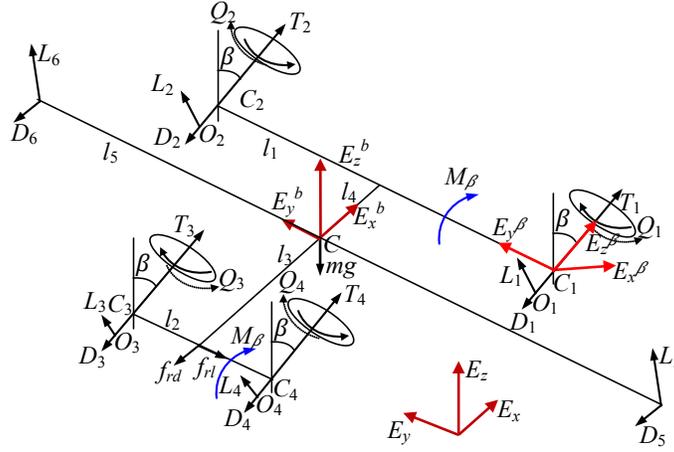

Fig. 4. Forces and torques of tilt-rotor aircraft

$p_b=(x_b, y_b, z_b)$ is the position of center of gravity relative to frame $\Gamma_g$. $(\psi,\theta,\phi)\in\Re^3$ describes the aircraft orientation expressed in the classical yaw, pitch and roll angles (Euler angles). We use $c_\theta$ for $\cos\theta$ and $s_\theta$ for $\sin\theta$. $R_{bg}$ is the transformation matrix representing the orientation of the rotorcraft from frame $\Gamma_b$ to $\Gamma_g$, i.e.,

$$R_{bg} = \begin{bmatrix} c_\theta c_\psi & c_\phi s_\psi + s_\phi s_\theta c_\psi & s_\phi s_\psi - c_\phi s_\theta c_\psi \\ -c_\theta s_\psi & c_\phi c_\psi - s_\phi s_\theta s_\psi & s_\phi c_\psi + c_\phi s_\theta s_\psi \\ s_\theta & -s_\phi c_\theta & c_\phi c_\theta \end{bmatrix} \tag{22}$$

$\beta$ is the tilt angle of rotors. $\Gamma_{\beta b}$ is the transformation matrix representing the orientation of the rotorcraft from frame $\Gamma_\beta$ to $\Gamma_b$, i.e.,

$$R_{\beta b} = \begin{bmatrix} c_\beta & 0 & s_\beta \\ 0 & 1 & 0 \\ -s_\beta & 0 & c_\beta \end{bmatrix} \tag{23}$$

**3.1.2 Aerodynamic analysis of large-size tilt rotor**

Algebraic expressions of the aerodynamic forces and torques are used to deduce the generalized external forces acting on the aircraft. In this subsection, using the blade element method [27, 28], the force and torque computations are presented, i.e., the following expressions are formulated, respectively: rotor thrust and blade pitch angle, drag torque and blade pitch angle. Moreover, the relationship between the drag torque and rotor thrust is obtained.

The airflow velocity vector in reference system $\Gamma_\beta$ which is located at the center of each rotor is

$$\vec{V}_\beta = \begin{bmatrix} V_{\beta X} & V_{\beta Y} & V_{\beta Z} \end{bmatrix}^T = R_{\beta b}^{-1} \vec{V}_b \tag{24}$$

where $\vec{V}_b$ is the airflow velocity vector in frame $\Gamma_b$. Let

$$\bar{v}_i = \frac{v_i}{\Omega R}, \bar{r} = \frac{r}{R}, \mu_x = \frac{V_{\beta X}}{\Omega R}, \mu_y = \frac{V_{\beta Y}}{\Omega R}, \mu_z = \frac{V_{\beta Z}}{\Omega R}, \mu_\beta = \frac{\dot{\beta}r}{\Omega R} \tag{25}$$

$$\bar{W}_x = \bar{r} + \mu_x \sin\gamma + \mu_y \cos\gamma, \; \bar{W}_y = \mu_x \cos\gamma + \mu_y \sin\gamma,$$
$$\bar{W}_z = \bar{v}_i + \mu_z + \mu_\beta \cos\gamma, \; \rho_\beta = \dot{\beta}/\Omega \tag{26}$$

where $v_i$ is the induced velocity of rotor, and $\Omega$ is the rotor speed.

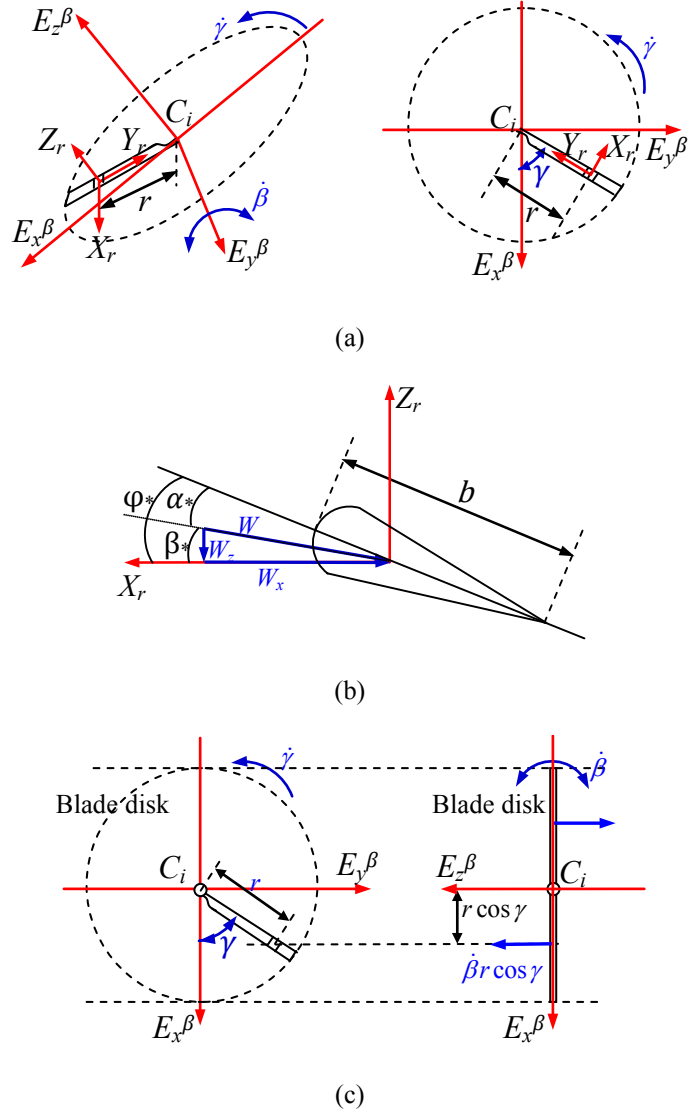

Fig. 5. Coordinates and forces for tilt-rotor: (a) Blade and blade element coordinates; (b) Forces on blade element; (c) Tilt dynamics on blade

**Assumption 1:** The following assumptions are taken into account in this development:

i) During mode transition, rotor speed $\Omega$ is constant such that the outputs of the four engines are all in rated power;

ii) The relative inflow angle at the blade element (See Figure 5(b)) will be $\beta^*=\arctan(W_z/W_x)\approx W_z/W_x$;

iii) Tilt rate $\dot{\beta}$ is far smaller than rotor speed $\Omega$, i.e.,

$$|\dot{\beta}| \ll \Omega = \dot{\gamma} \tag{27}$$

The blades without blade cyclic pitch are selected. And for the blade with linear torsion $\Delta\varphi_*$, the blade pitch angle $\varphi_*$ is expressed as

$$\varphi_* = \varphi_{7i} + \Delta\varphi_* (\bar{r} - 0.7) \tag{28}$$

where $\Delta\varphi_*=\varphi_1-\varphi_0$, and $\varphi_{7i}$ is pitch angle at $\bar{r}=0.7$. The chord length of blade with wingtip is constant, and we let $\bar{b}=\bar{b}_1$, where $\bar{b}=b/R$.

**Remark 1:** Assumption 1 is presented for the large-size aircrafts with turboshaft engines, which haven't variable-speed rotors. Small-size tilt-rotor aircrafts are usually electric powered. For an electric powered tilt-rotor aircraft, the rotor speeds are allowed to freely vary according to the desired level of thrust. However, the rotor speeds varying freely according to the desired level of thrust is unrealistic for current state-of-the art technology for conventional large-size rotorcrafts, i.e., the control of the thrust output is usually difficult to be achievable by changing the rotor speed. For the large-size aircrafts with turboshaft engines, the blade pitch angles of the rotors are regulated to vary according to the desired level of thrust.

In fact, using a variable velocity joint between the rotor and engine is a means to allow for speed changes of the rotor. By changing the transmission ratio of gearbox or belt wheels, the speed change of rotor is implemented. However, for the gearbox, speed range is limited in rated power, and it only has several speeds. Continuously variable transmission (CVT), also known as single-speed transmission, gearless transmission, variable pulley transmission, is a transmission in which the ratio of the rotational speeds of two shafts, as the input (or engine) shaft and output shaft, can be varied continuously within a given range, providing an infinite number of possible ratios. The CVT is allowed to select the relationship between the speed of the engine and the speed of the wheels within a continuous range, and the transmission action is smooth. However, it is not suitable for high maneuverability and large-torque load, and it offers approximately 88% efficiency. With respect to using a variable velocity joint, controlling the blade pitch angles of the rotors by rudders is very fast and accurate, and the rotor speeds are fixed in rated power.

Other than the Boeing A160 Hummingbird, all rotorcrafts to date feature fixed speed rotors (the V-22 is occasionally referred to as having a variable-speed rotor, but it really has only two speeds). Unlike most rotorcrafts, which feature turboshaft engines for requirements on high power output, the A160 has a piston engine that allows for variable-speed rotors. A turboshaft version of the A160 does exist, but it hasn't overcome the technical challenge of being variable-speed yet. Unless the intended vehicle does not have high power requirements (which would be unlikely considering that the tilt-rotor aircrafts are highly inefficient in hovering flight relative to pure helicopters like the A160), the vehicle would need to have turboshaft engines, and it is impossible to be assumed that control of the thrust output is achievable by changing the rotor speed. Besides, the downside to having variable-speed rotors is that a departure from the optimal rotor speed means that engine efficiency is reduced.

Therefore, in this paper, the control of the thrust output is achievable by changing the blade pitch angles of the rotors, and the rotor speeds are fixed in rated power. ∎

*Rotor thrust generation*

**Theorem 1:** For the tilt-rotor, which is satisfied with Assumption 1, the following relationship between rotor thrust coefficient and blade pitch angle holds:

$$\varphi_{7i}=\frac{\frac{3C_T\pi}{p\bar{b}_1 a_\infty}+\left(0.3\mu_x^2+0.3\mu_y^2-0.05\right)\Delta\varphi_*+\frac{3}{2}\left(\bar{v}_i+\mu_z\right)+\frac{3}{4}\mu_y\rho_\beta}{1+\frac{3}{2}\mu_x^2+\frac{3}{2}\mu_y^2} \tag{29}$$

Especially, if the sideslip can be ignored during mode transition, i.e., $\mu_y=0$, then

$$\varphi_{7i}=\left[\frac{3C_T\pi}{p\bar{b}_1 a_\infty}+\left(0.3\mu_x^2-0.05\right)\Delta\varphi_*+\frac{3}{2}\left(\bar{v}_i+\mu_z\right)\right]\bigg/\left(1+\frac{3}{2}\mu_x^2\right) \tag{30}$$

The proof of Theorem 1 is presented in Appendix.

**Remark 2 (Relationship between rotor thrust and blade pitch angle):** From Eq. (106) in Appendix and Eq. (1), the rotor thrust generated by each rotor can be written as

$$T_i = h_{i1}\varphi_{7i} + h_{i2} \tag{31}$$

where

$$h_{i1} = \rho\Omega^2 R^4 p\bar{b}_1 a_\infty (1 + 1.5\mu_x^2)/6$$

$$h_{i2} = \rho\Omega^2 R^4 p\bar{b}_1 a_\infty \{(0.05 - 0.3\mu_x^2)\Delta\varphi_* - 1.5(\bar{v}_i + \mu_z)\}/6 \tag{32}$$

and in the frame $\Gamma_\beta$, the thrust vector is described by

$$\vec{T}_i = \begin{bmatrix} 0 & 0 & T_i \end{bmatrix}^T \tag{33}$$

*Blade drag torque*

For the blade element drag torque, considering of Eq. (100) in Appendix and Fig. 5, we obtain

$$dQ_i = (dL\sin\beta_* + dD\cos\beta_*)r \approx (dL\beta_* + dD)r = 0.5\rho b a_\infty \left(W_x^2 C_d/a_\infty - W_z^2 + W_x W_z \varphi_*\right) r dr \tag{34}$$

Taking into account Eq. (34), the number of blades, the blade element conditions and the total contribution in one revolution, we can write:

$$Q_i = \frac{p}{2\pi}\int_0^{2\pi}\int_0^R \frac{dM_M}{dr}dr d\gamma = \frac{\rho p a_\infty}{4\pi}\int_0^{2\pi}\int_0^R \left(W_x^2 \frac{C_d}{a_\infty} - W_z^2 + W_x W_z \varphi_*\right) brdrd\gamma \tag{35}$$

The drag torque coefficient $C_{Q_i}$ is defined as

$$C_{Q_i} = Q_i / (0.5\rho\Omega^2 R^2 \pi R^2 R) \tag{36}$$

**Theorem 2 (Relationship between drag torque coefficient and blade pitch angle):** For the tilt-rotor, which is satisfied with Assumption 1, the following relationship between drag torque coefficient blade pitch angle holds:

$$C_{Q_i} = \bar{d}_{i1}\varphi_{7i} + \bar{d}_{i2} \tag{37}$$

where

$$\bar{d}_{i1} = \frac{p\bar{b}_1 a_{\infty 1}}{3\pi}(\bar{v}_i + \mu_z + \frac{1}{2}\mu_y \rho_\beta)$$

$$\bar{d}_{i2} = \frac{p\bar{b}_1 a_\infty}{\pi}\{\frac{C_{d0}}{4a_\infty}(1+\mu_x^2+\mu_y^2) + \frac{1}{60}(\bar{v}_i + \mu_z + \frac{1}{2}\mu_y \rho_\beta)\Delta\varphi_* - \frac{1}{2}(\bar{v}_i^2 + \mu_z^2 + 2\bar{v}_i\mu_z + \frac{1}{4}\rho_\beta^2)\} \tag{38}$$

The proof of Theorem 2 is presented in Appendix.

**Remark 3:** From Eqs. (36) and (37), the drag torque can be rewritten as

$$Q_i = d_{i1}\varphi_{7i} + d_{i2} \tag{39}$$

where $d_{i1} = 0.5\rho\Omega^2 R^2 \pi R^2 R \bar{d}_{i1}, d_{i2} = 0.5\rho\Omega^2 R^2 \pi R^2 R \bar{d}_{i2}$.

**Remark 4:** We suppose the sideslip can be ignored during mode transition, i.e., $\mu_y = 0$. Consequently, Eq. (38) becomes

$$\bar{d}_{i1} = \frac{p\bar{b}_1 a_{\infty 1}}{3\pi}(\bar{v}_i + \mu_z)$$

$$\bar{d}_{i2} = \frac{p\bar{b}_1 a_\infty}{\pi}\{\frac{C_{d0}}{4a_\infty}(1+\mu_x^2+\mu_y^2) + \frac{1}{60}(\bar{v}_i+\mu_z)\Delta\varphi_* - \frac{1}{2}(\bar{v}_i^2 + \mu_z^2 + 2\bar{v}_i\mu_z)\} \quad (40)$$

**Remark 5:** From Eqs. (31) and (39), we obtain the relationship between thrust $T_i$ and drag torque $Q_i$ as follow:

$$Q_i = d_{i1} T_i / h_{i1} + (d_{i2} - d_{i1} h_{i2} / h_{i1}) \quad (41)$$

Moreover, from Fig. 4, the sum of the thrusts generated by the quad rotors is $T = \sum_{i=1}^{4} T_i$.

### 3.1.3 Lift and drag forces generated by wings

In this subsection, the aerodynamics analysis is presented for free wings and fixed wings, and the effect of rotor tilting is considered.

*Aerodynamics of free wing*

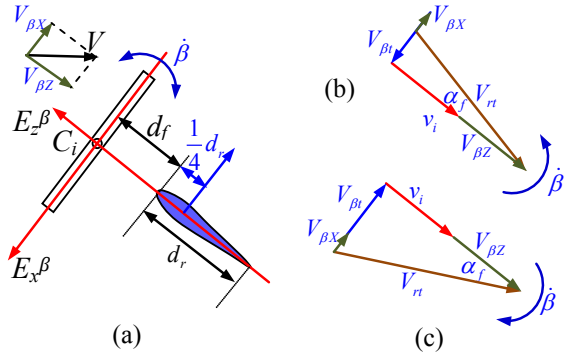

Fig. 6. Force on free wing: (a) tilt dynamics of free wing; (b) tilting from hover to forward flight; (c) tilting from forward flight to hover

For free wings, from Fig. 6, we obtain

$$V_{\beta t} = (d_f + 0.25 d_r)\dot{\beta}, \ V_{rt} = \left[(V_{\beta X} + V_{\beta t})^2 + (v_i + V_{\beta Z})^2\right]^{1/2}, \ \alpha_f = \arctan(V_{\beta X} + V_{\beta t})/(v_i + V_{\beta Z}) \quad (42)$$

where $V_{\beta t}$ is the airflow generated by rotor tilting, $V_{rt}$ is the resultant velocity at the free wing, and $\alpha_f$ is angle of attack for free wing, $d_f$ is the distance from the center of rotor to the leading edge of the free wing. Therefore, the lift and drag forces on free wing ($i=1,2,3,4$) are written, respectively, as

$$L_i = L_{i0} + \Delta L_i, \ L_{i0} = 0.5\rho S_{fi} C_{L_f} V_{rt}^2 \quad (43)$$

and

$$D_i = D_{i0} + \Delta D_i, \ D_{i0} = 0.5\rho S_{fi} C_{D_f} V_{rt}^2 \quad (44)$$

where $\Delta L_i$ and $\Delta D_i$ are the bounded uncertainties, and

$$C_{L_f} = C_f \alpha_f + C_r \delta_{fi}, \quad C_{D_f} = C_{D_{f0}} + C_{L_f}^2 / (\pi A_f e_f), \quad e_f = 1.78(1 - 0.045 A_f^{0.68}) - 0.46 \tag{45}$$

$S_{fi}$ is the area of the free wing, $C_f$ is the lift coefficient brought by angle of attack $\alpha_f$ to free wing, $C_{Df0}$ is the drag coefficient when the angle of attack $\alpha_f$ is equal to zero. $C_r$ is the lift coefficient brought out by the flap bias angle $\delta_{fi}$ on free wing. $e_f$ is the value of the Oswald's efficiency factor, and $A_f$ is the aspect ratio of free wing.

**Remark 6:** The case without Oswald's efficiency factor applies only to wings with elliptical lift distributions. It is possible to modify the lift coefficient expression slightly to make it apply any wing by using an Oswald's efficiency factor. The value of Oswald's efficiency factor is 1 for elliptical wings and between 0.5 and 1 for most common wings shapes. ∎

For the free wing, let the angle of attack with respect to the maximal lift coefficient be $\alpha_{max}$. In order to avoid the stall phenomenon, the following relation should hold:

$$|\alpha_f| = |\arctan((V_{\beta X} + V_{\beta t})/(v_i + V_{\beta Z}))| \leq \alpha_{max} \tag{46}$$

Therefore, from Eq. (46) and Figure 6(b) (i.e., tilting from hover to forward flight), for $d\beta/dt \geq 0$, we obtain

$$\dot{\beta} \leq [(v_i + V_{\beta Z}) \tan \alpha_{max} + V_{\beta X}]/(d_f + 0.25 d_r) \tag{47}$$

From Eq. (46) and Figure 6(c) (i.e., tilting from forward flight to hover), for $d\beta/dt < 0$, we obtain

$$-\dot{\beta} \leq [(v_i + V_{\beta Z}) \tan \alpha_{max} - V_{\beta X}]/(d_f + 0.25 d_r) \tag{48}$$

For the rear free wings (See Fig. 4), we select them as the controllers for the pitch dynamics when the tilt angle $\beta$ exceeds a given value. Therefore, we rewrite Eq. (43) as follow:

$$L_i = L_{i0}^1 + L_{i0}^2 + \Delta L_i, \quad i = 3, 4 \tag{49}$$

where $L_{i0}^1 = 0.5 \rho S_{fi} V_{rt}^2 C_r \delta_i$, $L_{i0}^2 = 0.5 \rho S_{fi} V_{rt}^2 C_f \alpha_f$.

The bias angle $\delta_3 = \delta_4 = \delta$ regulates the pitch dynamics.

*Aerodynamics of fixed wing*

The lift and drag forces on the fixed wings (See Fig. 4) are, respectively,

$$L_i = L_{i0} + \Delta L_i, \quad L_{i0} = 0.5 \rho S_{ri} C_{L_w} V_b^2, \quad i = 5, 6 \tag{50}$$

and

$$D_i = D_{i0} + \Delta D_i, \quad D_{i0} = 0.5 \rho S_{ri} C_{D_w} V_b^2 \tag{51}$$

where

$$C_{L_w} = C_{w0} + C_{w\alpha} \alpha + C_{w\delta} \delta_i, \quad C_{D_w} = C_{D_{w0}} + C_{L_w}^2 / (\pi A_w e_w), \quad e_w = 1.78(1 - 0.045 A_w^{0.68}) - 0.46 \tag{52}$$

$\Delta L_i$ and $\Delta D_i$ are the uncertainties affected by rotors tilting. $A_w$ is aspect ratio of fixed wing. $S_{ri}$ is the area of the left (right) fixed wing, $C_{w\alpha}$ is the lift coefficient brought by the angle of attack $\alpha$ of fixed wing, $C_{w0}$ is the lift coefficient when angle of attack $\alpha$ is equal to zero, $C_{Dw0}$ is the drag coefficient when the angle of attack $\alpha$ is equal to zero. $C_{w\delta}$ is the lift coefficient brought out by the normal flap bias angle $\delta_i$. $e_w$ is the value of the Oswald's efficiency factor.

### 3.1.4 Forces generated by vertical tail

From Fig. 4, the yaw and drag forces generated by vertical tail in frame $\Gamma_b$ are denoted by $\begin{bmatrix} 0 & -f_{rl} & 0 \end{bmatrix}^T$ and $\begin{bmatrix} -f_{rl} & 0 & 0 \end{bmatrix}^T$, respectively,.

### 3.1.5 Gyroscopic and counteractive moments

Tilting the rotors around the axes $l_1$ and $l_2$ creates gyroscopic moments which are perpendicular to these axes and to the spin axe ($E_z^\beta$). Indeed, these moments are defined by the cross product of the kinetic moments ($J_r \Omega_i E_z^\beta$) of the rotors and the tilt velocity vector. They are first expressed in the rotor frames as [31]

$$G_\beta = \sum_{i=1}^{4} (-1)^{i+1} J_r \Omega_i \dot{\beta} E_x^\beta \tag{53}$$

for the longitudinal tilting. In order to express these gyroscopic moments in the fixed body frame $\Gamma_b$, we should multiply the above equation by the rotational matrix $R_\beta$. We therefore find

$$G_b = \begin{bmatrix} G_{b1} & G_{b2} & G_{b3} \end{bmatrix}^T = R_\beta \sum_{i=1}^{4} (-1)^{i+1} J_r \Omega_i \dot{\beta} E_x^\beta \tag{54}$$

### 3.1.6 Thrust vectoring moment

Denote $C_1$, $C_2$, $C_3$ and $C_4$ the application points of the thrust $T_1$, $T_2$, $T_3$ and $T_4$, respectively. From Fig. 4, we can define $C_1C=(l_4,-l_1,0)$, $C_2C=(l_4,l_1,0)$, $C_3C=(-l_3,l_2,0)$ and $C_4C=(-l_3,-l_2,0)$ as the positional vectors expressed in $\Gamma_b$. Then, the moment exerted by the forces $T_i$ ($i=1,2,3,4$) on the airframe $\Gamma_b$ is

$$\tau_T = \sum_{i=1}^{4} \bar{T}_i \times C_i C \tag{55}$$

where $\bar{T}_i$ ($i=1,2,3,4$) is the thrust vector in frame $\Gamma_b$. After some computations, we obtain

$$\tau_T = \begin{bmatrix} (T_2 - T_1)l_1 c_\beta + (T_3 - T_4)l_2 c_\beta \\ (T_1 + T_2)l_4 c_\beta - (T_3 + T_4)l_3 c_\beta \\ -(T_2 - T_1)l_1 s_\beta - (T_3 - T_4)l_2 s_\beta \end{bmatrix} \tag{56}$$

### 3.1.7 Reactive torques

As the blades of the quad rotors rotate, they are subject to drag forces which produce torques around the aerodynamic center $C_i$. These moments act in opposite direction relative to $\Omega_i$. The reactive torque generated in free air by the rotors due to rotor drag is given by:

$$Q = \sum_{i=1}^{4} (-1)^{i+1} Q_i \tag{57}$$

After some computations, we obtain reactive torque vector in frame $\Gamma_b$ as follow:

$$Q_r = \begin{bmatrix} s_\beta \sum_{i=1}^{4} (-1)^{i+1} Q_i & 0 & c_\beta \sum_{i=1}^{4} (-1)^{i+1} Q_i \end{bmatrix}^T \tag{58}$$

### 3.1.8 Adverse reactionary moment

As described previously, this moment appears when forcing the rotors to tilt longitudinally. It depends especially on the propeller inertia $J_4$ and on tilt accelerations. This moment acts as a pitch moment and can be expressed in frame $\Gamma_b$ as follow:

$$\tau_\beta = J_4 \ddot{\beta} E_y^b \tag{59}$$

### 3.1.9 Moment generated by fixed and free wings

From Fig. 4, the moment generated by lift forces $L_5$ and $L_6$ of fixed wings in frame $\Gamma_b$ is obtained as follow:

$$\tau_\delta = \begin{bmatrix} \tau_{\delta_r} & \tau_{\delta p} & \tau_{\delta y} \end{bmatrix}^T = \begin{bmatrix} (L_6 - L_5)l_5 c_\alpha & 0 & (L_5 - L_6)l_5 s_\alpha \end{bmatrix}^T \quad (60)$$

Meanwhile, in frame $\Gamma_b$, the moment generated by free wings is given as follow:

$$\tau_f = \begin{bmatrix} \tau_{f\phi} \\ \tau_{f\theta} \\ \tau_{f\psi} \end{bmatrix} = \begin{bmatrix} (L_2 - L_1)l_1 s_\beta + (L_3 - L_4)l_2 s_\beta \\ (L_1 + L_2)l_4 s_\beta - (L_3 + L_4)l_3 s_\beta \\ (L_2 - L_1)l_1 c_\beta + (L_3 - L_4)l_2 c_\beta \end{bmatrix} \quad (61)$$

Because the moment generated by drag forces $D_5$ and $D_6$ of fixed wings is very small, it is taken as an uncertain vector defined in frame $\Gamma_b$, and it expressed as follow: $\tau_{fd}$=[ $\tau_{fd1}$ 0 $\tau_{fd3}$ ]$^T$.

Also, the moment generated by the drag forces $D_1, \cdots, D_4$ of free wings is very small, it is taken as an uncertain vector defined in frame $\Gamma_b$, and it is expressed as follow: $\tau_{xd}$=[ $\tau_{xd1}$ $\tau_{xd2}$ $\tau_{xd3}$ ]$^T$.

### 3.1.10 Moment generated by vertical tail

From Fig. 4, in frame $\Gamma_b$, the moment generated by vertical tail is given as follow: $\tau_t$=[0 0 $f_{rl}l_3$]$^T$.

### 3.1.11 The sum moment in the fixed-body frame

From the moments analysis above, the sum of the moment in the fixed-body frame is presented as follow: $\tau = \tau_t + Q_r + \tau_\beta + \tau_\delta + \tau_f + \tau_{xd} + \tau_{fd} + \tau_t$.

## 3.2 The motion equation of the aircraft

Aerodynamic interference effects between the different components are taken as the bounded uncertainties and included in the motion equations. The equations of motion written in terms of the centre of mass $C$ in the fixed axes of co-ordinate ($X, Y, Z$) are then

$$\ddot{x}_p = m^{-1}(u_p + \Gamma_p + \Delta_p) \quad (62)$$

$$\ddot{x}_a = J^{-1}(u_a + u_\beta + \Gamma_a + \Delta_a) \quad (63)$$

$$J_4\ddot{\beta} = -M_\beta \quad (64)$$

where, $J$=diag$\{J_1, J_2, J_3\}$; $J_1, J_2$, and $J_3$ are the three-axis moment of inertias; $J_4$ is the tilting moment of inertia; and
$x_p = \begin{bmatrix} X & Y & Z \end{bmatrix}^T, x_a = \begin{bmatrix} \phi & \theta & \psi \end{bmatrix}^T, \Gamma_p = \begin{bmatrix} \Gamma_{p1} & \Gamma_{p2} & \Gamma_{p3} \end{bmatrix}^T$,

$$u_p = \begin{bmatrix} u_{p1} \\ u_{p2} \\ u_{p3} \end{bmatrix} = \begin{bmatrix} (c_\theta c_\psi s_\beta + (s_\phi s_\psi - c_\phi s_\theta c_\psi)c_\beta) \\ (-c_\theta s_\psi s_\beta + (s_\phi c_\psi + c_\phi s_\theta s_\psi)c_\beta) \\ (s_\theta s_\beta + c_\phi c_\theta c_\beta) \end{bmatrix} T$$

$$\Gamma_{p1} = +(c_\theta c_\psi s_\alpha + (s_\phi s_\psi - c_\phi s_\theta c_\psi)c_\alpha)L_{5,6} + (-c_\theta c_\psi c_\alpha + (s_\phi s_\psi - c_\phi s_\theta c_\psi)s_\alpha)D_{5,6} + (-c_\theta c_\psi c_\beta + (s_\phi s_\psi - c_\phi s_\theta c_\psi)s_\beta)\sum_{i=1}^{4} L_{i0}$$

$$+ (-c_\theta c_\psi s_\beta - (s_\phi s_\psi - c_\phi s_\theta c_\psi)c_\beta)\sum_{i=1}^{4} D_{i0} + (-c_\phi s_\psi - s_\phi s_\theta c_\psi)f_{rl} - c_\theta c_\psi f_{rd}$$

$$\Gamma_{p2} = (-c_\theta s_\psi s_\beta + (s_\phi c_\psi + c_\phi s_\theta s_\psi)c_\beta)T + (-c_\theta s_\psi s_\alpha + (s_\phi c_\psi + c_\phi s_\theta s_\psi)c_\alpha)L_{5,6} + (c_\theta s_\psi c_\alpha + (s_\phi c_\psi + c_\phi s_\theta s_\psi)s_\alpha)D_{5,6}$$

$$+ (c_\theta s_\psi c_\beta + (s_\phi c_\psi + c_\phi s_\theta s_\psi)s_\beta)\sum_{i=1}^{4} L_{i0} + (c_\theta s_\psi s_\beta - (s_\phi c_\psi + c_\phi s_\theta s_\psi)c_\beta)\sum_{i=1}^{4} D_{i0} + (-c_\phi c_\psi + s_\phi s_\theta s_\psi)f_{rl} + c_\theta s_\psi f_{rd}$$

$$\Gamma_{p3} = (s_\theta s_\beta + c_\phi c_\theta c_\beta)T + (s_\theta s_\alpha + c_\phi c_\theta c_\alpha)L_{5,6} + (-s_\theta c_\alpha + c_\phi c_\theta s_\alpha)D_{5,6} + (-s_\theta c_\beta + c_\phi c_\theta s_\beta)\sum_{i=1}^{4} L_{i0} + (-s_\theta s_\beta - c_\phi c_\theta c_\beta)\sum_{i=1}^{4} D_{i0} + s_\phi c_\theta f_{rl} - s_\theta f_{rd} - mg$$

$$L_{5,6} = L_{50} + L_{60}, \ D_{5,6} = D_{50} + D_{60}, \ \Delta_p = \begin{bmatrix} \Delta_{p1} & \Delta_{p2} & \Delta_{p3} \end{bmatrix}^T = \begin{bmatrix} \overline{\Delta}_1 & \overline{\Delta}_2 & \overline{\Delta}_3 \end{bmatrix}^T \tag{65}$$

and

$$u_a = \begin{bmatrix} u_{a1} & u_{a2} & u_{a3} \end{bmatrix}^T = c_\beta u_{ac} + s_\beta u_{as}, \ u_{ac} = \begin{bmatrix} u_{11} \\ u_{21} \\ u_{31} \end{bmatrix} = \begin{bmatrix} (T_2 - T_1)l_1 + (T_3 - T_4)l_2 \\ (T_1 + T_2)l_4 - (T_3 + T_4)l_3 \\ \sum_{i=1}^{4}(-1)^{i+1}Q_i \end{bmatrix}, \ u_{as} = \begin{bmatrix} u_{31} \\ u_{22} \\ -u_{11} \end{bmatrix} = \begin{bmatrix} \sum_{i=1}^{4}(-1)^{i+1}Q_i \\ (L_{30}^1 + L_{40}^1)l_3 \\ -[(T_2 - T_1)l_1 + (T_3 - T_4)l_2] \end{bmatrix}$$

$$u_\beta = \begin{bmatrix} u_{\beta 1} & u_{\beta 2} & u_{\beta 3} \end{bmatrix}^T = \begin{bmatrix} 0 & M_\beta & 0 \end{bmatrix}^T, \ \Gamma_a = \begin{bmatrix} \Gamma_{a1} & \Gamma_{a2} & \Gamma_{a3} \end{bmatrix}^T$$

$$\Gamma_{a1} = (L_6 - L_5)l_5 c_\alpha + [(L_2 - L_1)l_1 + (L_3 - L_4)l_2]s_\beta, \ \Gamma_{a2} = (L_{10} + L_{20})l_4 s_\beta - (L_{30}^2 + L_{40}^2)l_3 s_\beta$$

$$\Gamma_{a3} = +(L_{50} - L_{60})l_5 s_\alpha + [(L_{20} - L_{10})l_1 + (L_{30} - L_{40})l_2]c_\beta + f_{rl} l_3$$

$$\Delta_a = \begin{bmatrix} \overline{\Delta}_4 & \overline{\Delta}_5 & \overline{\Delta}_6 \end{bmatrix}^T \tag{66}$$

with

$$\overline{\Delta}_4 = \tau_{xd1} + \tau_{fd1} + G_{b1} + \overline{\Delta}_{\phi u}, \ \overline{\Delta}_5 = +\tau_{xd2} + G_{b2} + \overline{\Delta}_{\theta u}, \ \overline{\Delta}_6 = \tau_{xd3} + \tau_{fd3} + G_{b3} + \overline{\Delta}_{\psi u} \tag{67}$$

$\overline{\Delta}_1, \overline{\Delta}_2, \overline{\Delta}_3, \overline{\Delta}_{\phi u}, \overline{\Delta}_{\theta u}$ and $\overline{\Delta}_{\psi u}$ are the bounded uncertainties.

### 3.3 Measurement sensors and actuators

*Position* (*X,Y*) can be obtained by Global Positioning System (GPS). The position data from GPS is sent to the processor in the aircraft for feedback control. *Altitude Z* is measured by an altimeter. *Attitude* ($\psi,\theta,\phi$) and *attitude rate* (d$\psi$/dt,d$\theta$/dt,d$\phi$/dt) can be obtained by an Inertial Measurement Unit (IMU). Most common attitude sensors are based on gyros. *Angle of attack* is measured by an angle of attack sensor. Rotor thrusts $T_1$, $T_2$, $T_3$ $T_4$, tilting torque $M_\beta$ and the bias angle $\delta_3=\delta_4=\delta$ of the rear two free wings are selected as the control actuators.

**Remark 7:** Angle of attack is quite simply the angle between the wing chord and the oncoming air that the wing is flying through. Angle of attack can be measured by an angle of attack sensor. For instance, Dynon Avionics was the manufacturer of affordable EFIS products (for instance, EFIS-D100) to offer a way to measure angle of attack. Through extensive wind tunnel testing, Dynon is able to offer an angle of attack (AOA) pilot probe that measures both angle of attack and airspeed when connected to any of the EFIS-based products. Some other products of angle of attack sensors: AMETEK Aerospace Angle-of-Attack (AOA) transducers, the 4239-01 angle of attack (AOA) sensors, AeroControlex, Thales angle of attack (AOA) sensor, *et al*.

## 4 OBSERVER DESIGN

In systems (62) and (63), $\dot{X}$, $\dot{Y}$, $\dot{Z}$, $\ddot{X}$, $\ddot{Y}$, $\ddot{Z}$, $\ddot{\phi}$, $\ddot{\theta}$ and $\ddot{\psi}$ are unknown. Moreover, $\Delta_p$ and $\Delta_a$ are uncertain vectors. In order to reconstruct the unknown states and the uncertainties of systems (62) and (63), we present the finite-time convergent observers, and a theorem is given as follow.

**Theorem 3:** For systems (62) and (63), the finite-time convergent observers are presented as follows:

i) For position dynamics (62), the finite-time convergent observer is designed as

$$\dot{\hat{x}}_{i1} = \hat{x}_{i2} - k_{p1}|\hat{x}_{i1} - y_{in}|^{2/3}\text{sign}(\hat{x}_{i1} - y_{pi})$$
$$\dot{\hat{x}}_{i2} = \hat{x}_{i3} + m^{-1}(u_{pi} + \Gamma_{pi}) - k_{p2}|\hat{x}_{i1} - y_{pi}|^{1/3}\text{sign}(\hat{x}_{i1} - y_{pi}) \quad (68)$$
$$\dot{\hat{x}}_{i3} = -k_{p3}\text{sign}(\hat{x}_{i1} - y_{pi})$$

where $i=1,2,3$, and $y_{p1}=X$, $y_{p2}=Y$, $y_{p3}=Z$, $s^3+k_{p1}s^2+k_{p2}s+k_{p3}$ is Hurwitz, and $k_{p3} > m^{-1}|d(\bar{\Delta}_i)/dt|$, $i=1,2,3$. Therefore, a time $t_s$ exists, for $t \geq t_s$, such that

$$\hat{x}_{11} = X, \hat{x}_{12} = \dot{X}, \hat{x}_{13} = m^{-1}\bar{\Delta}_1,$$
$$\hat{x}_{21} = Y, \hat{x}_{22} = \dot{Y}, \hat{x}_{23} = m^{-1}\bar{\Delta}_2, \quad (69)$$
$$\hat{x}_{31} = Z, \hat{x}_{32} = \dot{Z}, \hat{x}_{33} = m^{-1}\bar{\Delta}_3$$

ii) For attitude dynamics (63), the finite-time observer is designed as

$$\dot{\hat{z}}_{i1} = \hat{z}_{i2} + J_i^{-1}(u_{ai} + u_{\beta i} + \Gamma_{ai}) - k_{a1}|\hat{z}_{i1} - y_{ai}|^{1/2}\text{sign}(\hat{z}_{i1} - y_{ai}) \quad (70)$$
$$\dot{\hat{z}}_{i2} = -k_{a2}\text{sign}(\hat{z}_{i1} - y_{ai})$$

where $i=1,2,3$, and $y_{a1} = \dot{\phi}$, $y_{a2} = \dot{\theta}$, $y_{a3} = \dot{\psi}$, $y_{a3}=\psi$, $s^2+k_{a1}s+k_{a2}$ is Hurwitz; and $k_{a3} > J_i^{-1}|d(\bar{\Delta}_i)/dt|$, $i=4,5,6$. Therefore, a finite time $t_s>0$ exists, for $t \geq t_s$, such that

$$\hat{z}_{11} = \dot{\phi}, \hat{z}_{12} = J_1^{-1}\bar{\Delta}_4, \hat{z}_{21} = \dot{\theta}, \hat{z}_{22} = J_2^{-1}\bar{\Delta}_5, \hat{z}_{31} = \dot{\psi}, \hat{z}_{32} = J_3^{-1}\bar{\Delta}_6 \quad (71)$$

The proof of Theorem 3 is presented in Appendix.

## 5 CONTROLLER DESIGN

In this section, a control law is derived for the purpose of stabilization and trajectory tracking, controllers for the attitude, position and tilting dynamics are designed, respectively. By using the finite-time convergent observers proposed above, the unknown states and the generalized disturbance are reconstructed. Suppose the reference trajectory and its finite order derivatives are bounded, and they are generated directly. Therefore, the proposed control law and its performance are presented in the following theorems.

*Controller design of attitude dynamics*

**Theorem 4:** For attitude dynamics (63), to track the reference attitude $a_d=[\phi_d\ \theta_d\ \psi_d]^T$, if the controller is designed as

$$u_a = -J(k_1 e_a + k_2 \hat{\dot{e}}_a - \ddot{a}_d) - u_\beta - \Gamma_a - \hat{\Delta}_a \quad (72)$$

where $e_a = x_a - a_d$, $\dot{e}_a = \dot{x}_a - \dot{a}_d$, $\hat{\dot{e}}_a = \hat{z}_2 - \dot{a}_d$, $\hat{z}_2 = [\hat{z}_{12}\ \hat{z}_{22}\ \hat{z}_{32}]^T$, $\hat{\Delta}_a = [J_1\hat{z}_{13}\ J_2\hat{z}_{23}\ J_3\hat{z}_{33}]^T$, $k_1$ and $k_2$ are positive constants, the observations $\hat{\dot{e}}_a$ and $\hat{\Delta}_a$ are obtained from observer (70), then the closed-loop system rendering by controller (72) will converge asymptotically to the origin, i.e., the tracking error $e_a \to 0$ and $de_a/dt \to 0$ as $t \to \infty$.

The proof of Theorem 4 is presented in Appendix.

### Switching logic for controller $u_a$

During mode transition, the flight speed is small when tilt angle β is small, and the role of the fixed wings is weak. The pitch dynamics is regulated by the differential control of the quad rotor thrusts; the flight speed becomes large when tilt angle β is large, and the effect of the fixed wings is strong. Moreover, the differential control of the quad rotor thrusts is week for the pitch dynamics. Therefore, the rear free wings can be selected as controller of the pitch dynamics.

Based on the analysis above, we will allocate controller $u_a$ with a switching logic based on the selection of β. It is supposed that we have selected a tilt angle $\beta_w$, and at this angle the controllers switches. Angle $\beta_w$ will be decided later.

i) When $\beta \leq \beta_w$, from Eq. (66), let

$$u_{22} = (L^1_{30} + L^1_{40})l_3 = 0 \text{ where } \delta_3 = \delta_4 = 0 \tag{73}$$

Therefore, controller (72) can be rewritten as

$$u_a = M_c u_{ac} = -J(k_1 e_a + k_2 \hat{\dot{e}}_a - \ddot{a}_d) - u_\beta - \Gamma_a - \hat{\Delta}_a \tag{74}$$

where $u_{ac}$ has been defined in Eq. (66), and

$$M_c = \begin{bmatrix} c_\beta & 0 & s_\beta \\ 0 & c_\beta & 0 \\ -s_\beta & 0 & c_\beta \end{bmatrix}, |M_c| = c_\beta, M_c^{-1} = \begin{bmatrix} c_\beta & 0 & s_\beta \\ 0 & c_\beta^{-1} & 0 \\ -s_\beta & 0 & c_\beta \end{bmatrix} \tag{75}$$

Then we obtain

$$u_{ac} = -M_c^{-1}[J(k_1 e_a + k_2 \hat{\dot{e}}_a - \ddot{a}_d) - u_\beta - \Gamma_a - \hat{\Delta}_a] \tag{76}$$

ii) When $\beta > \beta_w$, from Eq. (66), let

$$u_{21} = (T_1 + T_2)l_4 - (T_3 + T_4)l_3 = 0 \tag{77}$$

and

$$u_{22} = (L^1_{30} + L^1_{40})l_3, \text{ where } \delta_3 = \delta_4 = \delta \tag{78}$$

Therefore, controller (72) can be rewritten as

$$u_a = M_s u_{as} = -J(k_1 e_a + k_2 \hat{\dot{e}}_a - \ddot{a}_d) - u_\beta - \Gamma_a - \hat{\Delta}_a \tag{79}$$

where $u_{as}$ has been defined in Eq. (66), and

$$M_s = \begin{bmatrix} s_\beta & 0 & -c_\beta \\ 0 & s_\beta & 0 \\ c_\beta & 0 & s_\beta \end{bmatrix}, |M_s| = s_\beta, M_s^{-1} = \begin{bmatrix} s_\beta & 0 & -c_\beta \\ 0 & s_\beta^{-1} & 0 \\ c_\beta & 0 & s_\beta \end{bmatrix} \tag{80}$$

Then we obtain

$$u_{as} = -M_s^{-1}[J(k_1 e_a + k_2 \hat{e}_a - \ddot{a}_d) - u_\beta - \Gamma_a - \hat{\Delta}_a] \quad (81)$$

### Selection of tilt angle $\beta_w$

We expect the matrixes $M_c$ and $M_s$ have the same gain during the controllers switching. Therefore, we obtain $c_{\beta w} = s_{\beta w}$. Thus $\beta_w = \pi/4$.

### Controller design for position dynamics

**Theorem 5:** For position dynamics (62), to track the reference trajectory $x_d = [X_d \ Y_d \ Z_d]^T$, if the controller is designed as

$$u_p = -m(k_3 e_p + k_4 \hat{e}_p) - \Gamma_p - \hat{\Delta}_p + m\ddot{x}_d \quad (82)$$

where $e_p = x_p - x_d$, $\dot{e}_p = \dot{x}_p - \dot{x}_d$, $\hat{e}_a = \hat{x}_2 - \dot{x}_d$, $\hat{x}_2 = [\hat{x}_{12} \ \hat{x}_{22} \ \hat{x}_{32}]^T$, $\hat{\Delta}_p = [m\hat{x}_{13} \ m\hat{x}_{23} \ m\hat{x}_{33}]^T$, $k_3$ and $k_3$ are positive constants, the observations $\hat{e}_p$ and $\hat{\Delta}_p$ are obtained from observer (68), then the closed-loop system rendering by controller (82) will converge asymptotically to the origin, i.e., the tracking error $e_p \to 0$ and $\dot{e}_p \to 0$ as $t \to \infty$.

The proof of Theorem 5 is presented in Appendix.

### Controller design for tilting dynamics

During the hover-to-level transition, we expect that the flight speed attains a sufficient large value such that the fixe wings can provide the sufficient lift force. The tilt angle β is required to be tilted from zero degree to π/2. On the other hand, during the level-to-hover transition, we expect the flight speed attain zero, and hover be implemented. The tilt angle β is required to be tilted from π/2 degree to zero. Thus, the bounded controllers should be designed to implement these tilting actions during mode transitions. In following, we firstly give the desired tilting actions for hover-to-level and level-to-hover transitions, respectively.

*i) Tilt for hover-to-level transition:* $(\beta, \dot{\beta}, \ddot{\beta})$ is designed as:

$$\ddot{\beta}_d = \begin{cases} M_t, & 0 \leq t \leq t_1 \\ -M_t, & t_1 < t \leq 2t_1 \\ 0, & t > 2t_1 \end{cases}, \quad \dot{\beta}_d = \begin{cases} M_t t, & 0 \leq t \leq t_1 \\ M_t t_1 - M_t(t - t_1), & t_1 < t \leq 2t_1 \\ 0, & t > 2t_1 \end{cases}$$

$$\beta_d = \begin{cases} 0.5 M_t t^2, & 0 \leq t \leq t_1 \\ 0.5 M_t t_1^2 + M_t t_1 (t - t_1) - 0.5 M_t (t - t_1)^2, & t_1 < t \leq 2t_1 \\ M_t t_1^2, & t > 2t_1 \end{cases} \quad (83)$$

Because $\beta_d$ is required to attain $\pi/2$ when the hover-to-level transition finishes, we require $M_t t_1^2 = \pi/2$. Then $M_t = \pi/(2t_1^2)$. When $t = t_1$, $d\beta_d/dt$ attains its maximal value. Therefore, the following relation should be satisfied:

$$\dot{\beta}_d(t_1) = M_t t_1 < \frac{(v_{id} + V_{\beta Zd}) \tan \alpha_{\max} + V_{\beta Xd}}{d_f + 0.25 d_r} \bigg|_{t=t_1} \quad (84)$$

where $V_{\beta Xd}$ and $V_{\beta Zd}$ are the projects on $E_x^\beta$ and $E_z^\beta$, respectively, on frame $\Gamma_\beta$ for the desired flight speed $\dot{x}_d = [\dot{X}_d \ \dot{Y}_d \ \dot{Z}_d]^T$. Therefore, we obtain

$$t_1 > 0.5\pi(d_f + 0.25d_r)\Big/[(v_i + V_{\beta Z})\tan\alpha_{\max} + V_{\beta X}]\Big|_{t=t_1} \quad (85)$$

*ii) Tilt for hover-to-level transition:* ($\beta, \dot\beta, \ddot\beta$) is designed as:

$$\ddot\beta_d = \begin{cases} -M_t, & 0 \le t \le t_1 \\ M_t, & t_1 < t \le 2t_1 \\ 0, & t > 2t_1 \end{cases}, \quad \dot\beta_d = \begin{cases} -M_t t, & 0 \le t \le t_1 \\ -M_t t_1 + M_t(t-t_1), & t_1 < t \le 2t_1 \\ 0, & t > 2t_1 \end{cases}$$

$$\beta_d = \begin{cases} 0.5\pi - 0.5 M_t t^2, & 0 \le t \le t_1 \\ 0.5\pi - 0.5 M_t t_1^2 - M_t t_1 (t-t_1) + 0.5 M_t (t-t_1)^2, & t_1 < t \le 2t_1 \\ 0.5\pi - M_t t_1^2, & t > 2t_1 \end{cases} \quad (86)$$

where $M_t$ is positive and constant, and $t_1$ is the switching time. Because $\beta_d$ is required to attain zero when the level-to-hover transition finishes, we require $\pi/2 - M_t t_1^2 = 0$. Then $M_t = \pi/(2t_1^2)$. When $t=t_1$, $d\beta_d/dt$ attains its maximal value. Therefore, the following relation should be satisfied:

$$-\dot\beta_d(t_1) = M_t t_1 < \frac{(v_{id} + V_{\beta Zd})\tan\alpha_{\max} - V_{\beta Xd}}{d_f + 0.25 d_r}\Big|_{t=t_1} \quad (87)$$

Therefore, we obtain

$$t_1 > 0.5\pi(d_f + 0.25d_r)\Big/[(v_{id} + V_{\beta Zd})\tan\alpha_{\max} - V_{\beta Xd}]\Big|_{t=t_1} \quad (88)$$

For tilting dynamics (64), let $e_\beta = \beta - \beta_d$, $\dot e_\beta = \dot\beta - \dot\beta_d$, the system error is $\ddot e_\beta = -J_4^{-1} M_\beta - \ddot\beta_d$. The controller $M_\beta$ can be selected as

$$M_\beta = J_4(k_5 e_\beta + k_6 \dot e_\beta - \ddot\beta_d) \quad (89)$$

where $k_5, k_6 > 0$ are constant. The closed-loop system is

$$\ddot e_\beta = -k_5 e_\beta - k_6 \dot e_\beta \quad (90)$$

Selecting the Lyapunov function be $V_a = k_5 e_\beta^T e_\beta / 2 + \dot e_\beta^T \dot e_\beta / 2$, we can obtain that $e_\beta \to 0$ and $\dot e_\beta \to 0$ as $t \to \infty$.

## 6   COMPUTATIONAL ANALYSIS AND SIMULATION EXPERIMENTS

In this section, simulation results of mode transitions are presented in order to observe the performance of the proposed aircraft model and control law. We consider two cases of hover-to-level and level-to-hove transitions. The parameters used for the aircraft model are given in Table 1. These values are based on the previous modeling, observer and control law design in this paper.

i) In the first simulation part, the hover-to-level transition is considered. The desired hover-to-level transition can be described as follow: The aircraft starts its mission in hover at the desired height $Z_d$=100m, and its desired starting position is ($X_d(0), Y_d(0)$) = (0,0); The forward flight speed changes from zero to 100m/s by tilting the rotors synchronously; During the mode transition, the height is kept invariant. The desired attitude angle is fixed at ($\phi_d, \theta_d, \psi_d$) = (0°, 5°, 0°). We take the following initial conditions ($X(0), Y(0), Z(0), \phi(0), \theta(0), \psi(0)$) = (1, -1, 102, 5°, 3°, -5°), ($T_1(0), T_2(0), T_3(0), T_4(0)$) = (8117N, 8117N, 8117N, 8117N). The uncertainties in the aircraft dynamics are assumed as follows:

$$\Delta_p = \begin{bmatrix} \Delta_{p1} \\ \Delta_{p2} \\ \Delta_{p3} \end{bmatrix} = \begin{bmatrix} \overline{\Delta}_1 \\ \overline{\Delta}_2 \\ \overline{\Delta}_3 \end{bmatrix} = 50 \begin{bmatrix} 2\exp(-2t)\sin(3t) + \exp(-t)\cos(t) \\ \exp(-t)\sin(3t) + 2\exp(-0.5t)\cos(t) \\ 0.5\exp(-t)\sin(3t) + 3\exp(-2t)\cos(t) \end{bmatrix}, \quad \Delta_a = \begin{bmatrix} \overline{\Delta}_4 \\ \overline{\Delta}_5 \\ \overline{\Delta}_6 \end{bmatrix} = 20 \begin{bmatrix} 0.5\exp(-2t)\sin(3t) + 0.8\exp(-t)\cos(t) \\ 0.5\exp(-t)\sin(3t) + 0.5\exp(-0.5t)\cos(t) \\ 2\exp(-2t)\sin(3t) + 0.5\exp(-t)\cos(t) \end{bmatrix}$$

The results in Fig. 7, obtained by considering the complete mode of the aircraft, illustrate the performance of hover-to-level transition: Fig. 7(a) presents the position in X direction; Fig. 7(b) describes the velocity in X direction; Fig. 7(c) describes the position in Y direction; Fig. 7(d) presents the velocity in Y direction; Fig. 7(e) describes the position in Z direction; Fig. 7(f) describes the velocity in Z direction; Figs. 7(g) and 7(h) describe the roll angle and roll rate, respectively; Figs. 7(i) and 7(j) describe the pitch angle and pitch rate, respectively; Figs. 7(k) and (*l*) describe the yaw angle and yaw rate, respectively; Figs. 7(m)-7(o) show the tilt angle, tilt rate and tilt acceleration, respectively; Fig. 7(p) shows the four thrusts generated by the quad rotors.

ii) A second part, we ran simulations for level-to-hover transition. The desired level-to-hover transition can be described as follow: The aircraft starts its mission in forward flight with the flight speed $V$=100m/s at the desired height $Z_d$=100m, and its desired starting position is $(X_d(0), Y_d(0))$ = (0,0); The forward flight speed changes from 100m/s to zero by tilting the rotors synchronously; During the mode transition, the height is kept invariant. The desired attitude angle is fixed at $(\phi_d, \theta_d, \psi_d)$ = (0°, 0°, 0°). We take the following initial conditions $(X(0), Y(0), Z(0), \phi(0), \theta(0), \psi(0))$ = (1, -1, 102, 5°, 3°, -5°), $(T_1(0), T_2(0), T_3(0), T_4(0))$ = (1917N, 1917N, 1917, 1917N). The uncertainties in the aircraft dynamics are the same as in hover-to-level transition. The results in Fig. 8 illustrate the performance of the level-to-hover transition.

| parameter | value | parameter | value |
|---|---|---|---|
| $m$ | 3313kg | $g$ | 9.8m/s$^2$ |
| $m_e$ | 195kg | $\rho_e$ | 6 |
| $\rho_{mT}$ | $\sqrt{2}$ | $DL_{QT}$ | 60kg/m$^2$ |
| $R$ | 2.0966m | $p$ | 4 |
| $\rho_b$ | 13 | $\sigma$ | 0.1 |
| $S_w$ | 43.7500m$^2$ | $S_{front}$ | 12.3458m$^2$ |
| $S_{ft}$ | 4.3795m$^2$ | $l_1$ | 4.0900m |
| $l_2$ | 2.8050m | $l_3$ | 5.6800m |
| $l_4$ | 3.4900m | $b$ | 0.1613m |
| $a_\infty$ | 0.012 | $d_f$ | 2.0000m |
| $d_r$ | 1.0445m | $S_{ft}$ | 4.3795m$^2$ |
| $C_{Df0}$ | 0.008 | $C_r$ | 0.15 |
| $C_f$ | 0.5 | $A_f$ | 4 |
| $\alpha_{max}$ | 25° | $A_w$ | 12 |
| $S_{ri}$ | 21.8750m² | $C_{w0}$ | 0.32 |
| $C_{w\alpha}$ | 0.5 | $C_{Dw0}$ | 0.008 |
| $C_{w\delta}$ | 0.15 | $J_r$ | 8.5m²kg |
| $J_1$ | 220m²kg | $J_2$ | 220m²kg |
| $J_3$ | 400m²kg | $J_4$ | 50m²kg |
| $\theta_d$ | 5° | $\rho$ | 1.225kg/m³ |
| $dx_d/dt$ | 100m/s | $\beta_w$ | 45° |
| $k_{p1}$ | 6 | $k_{p2}$ | 11 |
| $k_{p3}$ | 6 | $k_{a1}$ | 6 |
| $k_{a2}$ | 11 | $k_{a3}$ | 6 |
| $k_1$ | 5 | $k_2$ | 5 |
| $k_3$ | 2.5 | $k_4$ | 4.5 |
| $k_5$ | 2.63 | $k_6$ | 4.55 |
| $t_1$ | 5s | $\Delta\varphi_*$ | -7° |

Table 1 Parameters of aircraft model and flight

Although the uncertainties exist in the dynamics equations of the tilt-rotor aircraft, the presented controller based on the finite-time convergent observer drive the aircraft to the desired trajectories during the two mode transitions. We can find out that the thrust (1917$N$) generated by each rotor during forward flight mode is far smaller than that (8117$N$) in hover. Therefore, under the same cruise speed, the presented tilt-rotor aircraft can save much energy than the conventional helicopters. It can increase the cruise time and flying range. Moreover, the computational analysis and simulations exhibit the agile maneuverability of the presented tilt-rotor aircraft with the simple control algorithm.

## 7    CONCLUSION

In this paper, a novel 6-DOF model of large-size tilt-rotor aircraft is presented. Not only the proposed aircraft can hover, take off and land vertically, but also the high-speed forward flight can be implemented. The aerodynamics of mode transition with rotors tilting is analyzed. During mode transitions, using the blade element method, the mathematical modeling is proposed for the following expressions: rotor thrust and blade pitch angle, drag torque and blade pitch angle. Furthermore, using the Lyapunov function method, a finite-time convergent observer is designed to reconstruct the unknown states and uncertainties in the aircraft system. Finally, the presented switched logic controller based on the finite-time convergent observer can drive the aircraft to implement mode transitions. Although the uncertainties exist in the dynamics of the tilt-rotor aircraft during mode transitions, the strong stability and agile maneuverability are exhibited for the presented tilt-rotor aircraft. Our future work is to implement a prototype of the presented tilt-rotor aircraft.

# 8 APPENDIX

**Proof of Theorem 1:**

The mass flow rate, $\dot{m}$, through the actuator disk is $\dot{m} = \rho A U$, where $U$ is the resultant velocity at the disk and is given by $U = \sqrt{V_{\beta X}^2 + V_{\beta Y}^2 + (V_{\beta Z} + v_i)^2}$. The application of the conservation of momentum in a direction normal to the disk gives

$$T_i = 2\rho A v_i \sqrt{V_{\beta X}^2 + V_{\beta Y}^2 + V_{\beta Z}^2 + v_i^2 + 2V_{\beta Z} v_i} \tag{91}$$

Therefore,

$$v_i^4 + 2V_{\beta Z} v_i^3 + (V_{\beta X}^2 + V_{\beta Y}^2 + V_{\beta Z}^2) v_i^2 = \left(T_i / (2\rho \pi R^2)\right)^2 \tag{92}$$

For Eq. (92), Newton-Raphson iteration method can be adopted to obtain $v_i$. Especially, the sideslip can be ignored during mode transition, i.e., $V_{\beta Y}=0$. Therefore, Eq. (92) can be rewritten as

$$v_i^4 + 2V_{\beta Z} v_i^3 + (V_{\beta X}^2 + V_{\beta Z}^2) v_i^2 = \left(T_i / (2\rho \pi R_{4r}^2)\right)^2 \tag{93}$$

Normalize $V_{\beta X}$, $V_{\beta Z}$ and $v_i$ by the mean induced velocity at hover $v_h$

$$v_h = \sqrt{T_i / (2\rho \pi R^2)}, V_{\beta Xh} = V_{\beta X}/v_h, V_{\beta Zh} = V_{\beta Z}/v_h, v_{ih} = v_i/v_h \tag{94}$$

For rotor flows outside the vortex-ring state, $(2V_{\beta Zh} + 3)^2 + V_{\beta Xh}^2 > 1$, the normalized ideal induced velocity can be determined from the momentum theory [28, 29]

$$v_{ih}^4 + 2V_{\beta Zh} v_{ih}^3 + (V_{\beta Xh}^2 + V_{\beta Zh}^2) v_{ih}^2 = 1 \tag{95}$$

Inside the vortex-ring state $(2V_{\beta Zh} + 3)^2 + V_{\beta Xh}^2 \leq 1$, the induced velocity can be determined from an approximation proposed by Johnson [30]:

$$v_{ih} = V_{\beta Zh}(0.373 V_{\beta Zh}^2 + 0.598 V_{\beta Xh}^2 - 1.991) \tag{96}$$

In forward flight, the blade element velocity components are periodic at the rotor rotational frequency. As for the hover case, there is an in-plane velocity component because of blade rotation about the rotor shaft, but now there is a further free-stream (translational) part and tilt rotor velocity such that $W_x = r\Omega + V_{\beta X}\sin\gamma + V_{\beta Y}\cos\gamma$, and the velocity perpendicular to the disk can be written as $W_z = v_i + V_{\beta Z} + (d\beta/dt) r \cos\gamma$. Therefore, $W = \sqrt{W_x^2 + W_z^2} \approx W_x$. The resultant incremental lift $dL$ per unit span on the blade element is given by

$$dL = 0.5\rho W^2 b C_l dr = 0.5\rho b a_\infty \left(\varphi_* W_x^2 - W_x W_z\right) dr \tag{97}$$

and the incremental drag is

$$dD = 0.5\rho W^2 bC_d dr \approx 0.5\rho W_x^2 bC_d dr \tag{98}$$

where $C_l$ is the lift coefficient, $C_d$ is the drag coefficient, $a_\infty$ is the slope of the lift curve, $b$ is the blade chord, $\alpha_*$ is the angle of attack of the blade element, and $dr$ is the incremental radial distance. Therefore, for a blade differential element, the incremental lift force $dT_i$ is

$$dT_i = dL\cos\beta_* - dD\sin\beta_* \tag{99}$$

and the drag torque gives

$$dQ_i = (dL\sin\beta_* + dD\cos\beta_*)r \tag{100}$$

The relative inflow angle at the blade section is $\cos\beta_* = W_x/W$, $\sin\beta_* = W_z/W$. Therefore,

$$dT_i = dL W_x/W - dD W_z/W \tag{101}$$

$$dQ_i = (dL W_z/W + dD W_x/W)r \tag{102}$$

The total trust of one rotor is equal the number of blades ($p$) times the average lift per bade:

$$T_i = p(2\pi)^{-1}\int_0^{2\pi}\int_0^R (dT/dr)dr d\gamma \tag{103}$$

The second term in (101) is very tiny with respect to the first term, thus, we obtain $dT_i \approx dL W_x/W \approx dL$. Therefore,

$$T_i = \frac{p}{2\pi}\int_0^{2\pi}\int_0^R \frac{dL}{dr}dr d\gamma = \frac{p}{2\pi}\frac{\rho}{2}a_\infty\int_0^{2\pi}\int_0^R \left(\varphi_* W_x^2 - W_x W_z\right)b dr d\gamma \tag{104}$$

Then from Eqs. (1), (25), (26) and (104), we obtain

$$C_{T_i} = \frac{p}{2\pi}\frac{1}{\pi}a_\infty\int_0^{2\pi}\int_0^1 \left(\varphi_* \overline{W}_x^2 - \overline{W}_x \overline{W}_z\right)\bar{b} d\bar{r} d\gamma$$
$$= \frac{1}{3}\frac{p}{\pi}\bar{b}_1 a_\infty\left\{\varphi_{7i}\left(1+\frac{3}{2}\mu_x^2 + \frac{3}{2}\mu_y^2\right) + \left(0.05 - 0.3\mu_x^2 - 0.3\mu_y^2\right)\Delta\varphi_* - \frac{3}{2}(\bar{v}_i + \mu_z) - \frac{3}{4}\mu_y \rho_\beta\right\} \tag{105}$$

From Eq. (105), $\varphi_{7i}$ can be carried out shown in Eq. (29). Especially, we suppose the sideslip can be ignored during mode transition, i.e., $V_{\beta Y}=0$. Therefore, we obtain $\mu_y=0$. Consequently, we obtain

$$C_{T_i} = \frac{1}{3}\frac{p}{\pi}\bar{b}_1 a_\infty\left\{\varphi_{7i}\left(1+\frac{3}{2}\mu_x^2\right) + \left(0.05 - 0.3\mu_x^2\right)\Delta\varphi_* - \frac{3}{2}(\bar{v}_i + \mu_z)\right\} \tag{106}$$

it follows that

$$\frac{3 C_{T_i} \pi}{p \bar{b}_1 a_\infty} = \varphi_{7i}\left(1+\frac{3}{2}\mu_x^2\right) + \left(0.05 - 0.3\mu_x^2\right)\Delta\varphi_* - \frac{3}{2}(\bar{v}_i + \mu_z) \tag{107}$$

Then,

$$\varphi_{7i}\left(1+\frac{3}{2}\mu_x^2\right) = \frac{3 C_{T_i}\pi}{p\bar{b}_1 a_\infty} + \left(0.3\mu_x^2 - 0.05\right)\Delta\varphi_* + \frac{3}{2}(\bar{v}_i + \mu_z) \tag{108}$$

and Eq. (30) holds. This concludes the proof. ∎

**Proof of Theorem 2:**

From Eqs. (26), (35) and (36), we obtain

$$C_{Q_i} = \frac{pa_\infty}{2\pi} \frac{1}{\pi} \int_0^{2\pi} \int_0^1 \left( \overline{W}_x^2 \frac{C_d}{a_\infty} - \overline{W}_z^2 + \overline{W}_x \overline{W}_z \varphi_* \right) \bar{b} \bar{r} d\bar{r} d\gamma \qquad (109)$$

The section profile drag coefficient, $C_d$, is assumed to be constant $C_{d0}$, then after integrations we obtain Eqs. (37) and (38). This concludes the proof. ∎

**Proof of Theorem 3:**

1) For position dynamics, let

$$\begin{aligned}
\hat{e}_{11} &= \hat{x}_{11} - X, \hat{e}_{12} = \hat{x}_{12} - \dot{X}, \hat{e}_{13} = \hat{x}_{13} - m^{-1}\overline{\Delta}_1 \\
\hat{e}_{21} &= \hat{x}_{21} - Y, \hat{e}_{22} = \hat{x}_{22} - \dot{Y}, \hat{e}_{23} = \hat{x}_{23} - m^{-1}\overline{\Delta}_2 \\
\hat{e}_{31} &= \hat{x}_{31} - Z, \hat{e}_{32} = \hat{x}_{32} - \dot{Z}, \hat{e}_{33} = \hat{x}_{33} - m^{-1}\overline{\Delta}_3
\end{aligned} \qquad (110)$$

The system error between (68) and (62) is

$$\begin{aligned}
\dot{\hat{e}}_{i1} &= \hat{e}_{i2} - k_{p1}|\hat{e}_{i1}|^{2/3}\operatorname{sign}(\hat{e}_{i1}) \\
\dot{\hat{e}}_{i2} &= \hat{e}_{i3} - k_{p2}|\hat{e}_{i1}|^{1/3}\operatorname{sign}(\hat{e}_{i1}) \\
\dot{\hat{e}}_{i3} &= -k_{p3}\operatorname{sign}(\hat{e}_{i1}) - \delta\overline{\Delta}_i
\end{aligned} \qquad (111)$$

where, $\delta\overline{\Delta}_i = d(\overline{\Delta}_i)/dt$, $i = 1,2,3$.

Select the Lyapunov function be

$$V_i = \xi^T P_p \xi \qquad (112)$$

where $\xi = \begin{bmatrix} |\hat{e}_{i1}|^{1/3}\operatorname{sign}(\hat{e}_{i1}) & \hat{e}_{i2} & \hat{e}_{i3} \end{bmatrix}^T$, and $P_p$ is a positive definite and symmetrical matrix with the following form:

$$P_p = \frac{1}{2}\begin{bmatrix} 2k_{p3} + k_{p1}^2 + k_{p2}^2 & -k_{p1} & -k_{p2} \\ -k_{p1} & 2 & 0 \\ -k_{p2} & 0 & 2 \end{bmatrix} \qquad (113)$$

The Lyapunov function $V_i$ is to study the stability of various differential equations and systems. Differentiating $V_i$ with respect to time yields

$$\dot{V}_i \leq -\left[ c_p / \left(\lambda_{\max}\{P_p\}\right)^{1/2} \right] V_i^{1/2} \qquad (114)$$

where $c_p$ is a positive constant. From the definition of finite-time stability [32, 33], there exists a time $t_s>0$, for $t \geq t_s$, the system (111) is finite-time convergent. Alternatively, the convergent results of system (111) can be proved by high-order sliding mode theory [34].

2) For attitude dynamics, let

$$\hat{e}_{41} = \hat{z}_{11} - \dot{\phi}, \hat{e}_{42} = \hat{z}_{12} - J_1^{-1}\overline{\Delta}_4$$
$$\hat{e}_{51} = \hat{z}_{21} - \dot{\theta}, \hat{e}_{52} = \hat{z}_{22} - J_2^{-1}\overline{\Delta}_5 \quad (115)$$
$$\hat{e}_{61} = \hat{z}_{31} - \dot{\psi}, \hat{e}_{62} = \hat{z}_{32} - J_3^{-1}\overline{\Delta}_6$$

The system error between (70) and (63) is

$$\dot{\hat{e}}_{i1} = e_{i2} - k_{a1}|\hat{e}_{i1}|^{1/2}\,\text{sign}(\hat{e}_{i1})$$
$$\dot{\hat{e}}_{i2} = -k_{a2}\text{sign}(\hat{e}_{i1}) - \delta\overline{\Delta}_i \quad (116)$$

where $\delta\overline{\Delta}_i = d(\overline{\Delta}_i)/dt$, $i = 4,5,6$. Select the Lyapunov function be

$$V_i = \varsigma^{\text{T}} P_a \varsigma \quad (117)$$

where $\varsigma = \left[|\hat{e}_{i1}|^{\frac{1}{2}}\,\text{sgn}(\hat{e}_{i1}) \quad \hat{e}_{i2}\right]^{\text{T}}$, and $P_a$ is a positive definite and symmetrical matrix with the following form:

$$P_a = \frac{1}{2}\begin{bmatrix} 4k_{a2} + k_{a1}^2 & -k_{a1} \\ -k_{a1} & 2 \end{bmatrix} \quad (118)$$

Differentiating $V_i$ with respect to time yields

$$\dot{V}_i \leq -\left[c_a\big/(\lambda_{\min}\{P\})^{1/2}\right]V_i^{1/2} \quad (119)$$

where $c_a$ is a positive constant. Therefore, the system error (116) is finite-time convergent. This concludes the proof. ∎

**Proof of Theorem 4:**

In the light of Theorem 3, for $t \geq t_s$, the observation signals $\hat{z}_2 = \dot{x}_a$, $\hat{\Delta}_a = \Delta_a$. Considering controller (72), the closed-loop system is

$$\ddot{e}_a = -k_1 e_a - k_2 \hat{\dot{e}}_a + J^{-1}(\Delta_a - \hat{\Delta}_a) \quad (120)$$

Eq. (120) is written as

$$\ddot{e}_a = -k_1 e_a - k_2 \dot{e}_a + k_2(\dot{x}_a - \hat{z}_2) + J^{-1}(\Delta_a - \hat{\Delta}_a) \quad (121)$$

Therefore, for $t \geq t_s$, the system error can be written as

$$\ddot{e}_a = -k_1 e_a - k_2 \dot{e}_a \quad (122)$$

Selecting the Lyapunov function be

$$V_a = k_1 e_a^{\text{T}} e_a/2 + \dot{e}_a^{\text{T}} \dot{e}_a/2 \quad (123)$$

we can obtain that $e_a \to 0$ and $\dot{e}_a \to 0$ as $t \to \infty$. This concludes the proof. ∎

**Proof of Theorem 5:**

In the light of Theorem 3, for $t \geq t_s$, the observations $\hat{x}_2 = \dot{x}_p$, $\hat{\Delta}_p = \Delta_p$ hold. Considering controller (82), the closed-loop system is

$$\ddot{e}_p = -k_3 e_p - k_4 \hat{\dot{e}}_p + m^{-1}(\Delta_p - \hat{\Delta}_p) \quad (124)$$

We rewrite Eq. (124) as

$$\ddot{e}_p = -k_3 e_p - k_4 \dot{e}_p + k_4(\dot{x}_d - \hat{x}_2) + m^{-1}(\Delta_p - \hat{\Delta}_p) \tag{125}$$

For $t \geq t_s$, the system error can be written as $\ddot{e}_p = -k_3 e_p - k_4 \dot{e}_p$, Selecting the Lyapunov function be $V_p = k_3 e_p^T e_p / 2 + \dot{e}_p^T \dot{e}_p / 2$, we can obtain that $e_p \to 0$ and $\dot{e}_p \to 0$ as $t \to \infty$. This concludes the proof. ∎

## 9 LIST OF SYMBOLS

$m_e$ weight of each engine

$T_e$ rotor thrust under rated power

$\rho_e$ thrust-to-weight ratio of engine

$m$ gross weight of aircraft

$T_i$ rotor thrust

$C_{Ti}$ rotor thrust coefficient

$\Omega_i$ rotor speed

$\rho$ air density

$\sigma$ rotor solidity

$p$ blade number

$R$ rotor radius

$\rho_b$ blade span-chord ratio

$A$ area of rotor disk

$g$ gravity acceleration

$C_{d0}$ constant profile drag coefficient

$\kappa$ induced power factor

$b$ chord length of blade

$PL$ power loading of rotor

$DL$ disk loading of rotor

$R_{2r}$ rotor radius of aircraft with two rotors

$R_e$ Reynolds number

$M$ Mach number

$DL_{QT}$ Desired disk loading of rotor

$A_w$ span-chord ratio of fixed wing

$WL_w$ desired wing loading of fixed wing

$S_w$ wing area of fixed wing

$L_w$ wing span of fixed wing

$C_w$ chord length of fixed wing

$S_{front}$ area of front fixed wing root

$S_{fi}$ area of free wing

$\Gamma_g$ right handed inertial frame

$\Gamma_b$ frame attached to the aircraft's fuselage

$\Gamma_\beta$ right handed inertial frame of the tiltrotor

$\psi$ yaw angle

$\theta$ pitch angle

$\phi$ roll angle

$R_{bg}$ transformation matrix for $\Gamma_b$ to $\Gamma_g$

$R_{\beta b}$ transformation matrix for $\Gamma_\beta$ to $\Gamma_b$

$V_{\beta X}$ air relative velocity on axis $E_x^\beta$

$V_{\beta Y}$ air relative velocity on axis $E_y^\beta$

$V_{\beta Z}$ air relative velocity on axis $E_z^\beta$

$V_b$ air relative velocity with respect to frame $\Gamma_b$

$v_i$ induced velocity of rotor disk

$v_h$ induced velocity in hover

$r$ length from rotor center to blade element

$\beta_*$ relative inflow angle at blade element

$\alpha_*$ angle of attack of blade element

$\varphi_*$ blade pitch angle

$\varphi_{7i}$ pitch angle at $r/R=0.7$

$\Delta\varphi_*$ linear torsion of blade

$U$ resultant velocity at disk

$dL$ lift per unit span on blade element

$dD$ drag per unit span on blade element

$C_l$ lift coefficient of blade

$C_d$ drag coefficient of blade

$dr$ incremental radial distance of blade

$dT_i$ incremental lift of rotor

$dQ_i$ incremental drag torque of rotor

$Q_i$ drag torque of rotor

$C_{Qi}$ drag torque coefficient of rotor

$T$ sum of rotor thrusts

$V_{\beta t}$ airflow generated by rotor tilting

$V_{rt}$ resultant velocity at the free wing

$\alpha_f$ angle of attack for free wing

$d_f$ distance between rotor center to leading edge of free wing

$L_i$ ($i=1,2,3,4$) lift force generated by free wing

$D_i$ ($i=1,2,3,4$) lift force generated by free wing

$\Delta L_i$ ($i=1,2,3,4$) uncertain lift force for free wing

$\Delta D_i$ ($i=1,2,3,4$) uncertain drag force for free wing

$S_{fi}$ area of free wing

$C_f$ lift coefficient brought by angle of attack $\alpha_f$ to free wing

$C_{Df0}$ drag coefficient when angle of attack $\alpha_f$ is equal to zero

$\delta_i$ ($i=1,2,3,4$) flap bias angle of free wing

$C_r$ lift coefficient brought out by flap bias angle $\delta_i$

$e_f$ value of the Oswald's Efficiency Factor for free wing

$A_f$ span-chord ratio of free wing

$\alpha_{\max}$ angle of attack with respect to maximal lift coefficient of free wing

$L_i$ ($i=5,6$) lift force generated by fixed wing

$D_i$ ($i=5,6$) lift force generated by fixed wing

$\Delta L_i$ ($i$=5,6) uncertain lift force for fixed wing

$\Delta D_i$ ($i$=5,6) uncertain drag force for fixed wing

$\alpha$ angle of attack for fixed wing

$S_{ri}$ area of the left (right) fixed wing

$C_{w\alpha}$ lift coefficient brought by angle of attack $\alpha$ to fixed wing

$C_{w0}$ lift coefficient when angle of attack $\alpha$ is equal to zero

$C_{Dw0}$ drag coefficient when angle of attack $\alpha$ is equal to zero

$\delta_i$ ($i$=5,6) flap bias angle of fixed wing

$C_{w\delta}$ lift coefficient brought out by flap bias angle $\delta_i$

$e_w$ value of the Oswald's Efficiency Factor for fixed wing

$f_{rl}$ yaw force generated by vertical tail

$f_{rd}$ drag force generated by vertical tail

$G_\beta$ gyroscopic moment for tilting rotors in frame $\Gamma_\beta$

$G_b$ gyroscopic moment for tilting rotors in frame $\Gamma_b$

$\tau_T$ thrust vectoring moment in frame $\Gamma_b$

$C_i$ ($i$=1,2,3,4) center of rotor

$Q$ sum of drag torques of rotors in frame $\Gamma_\beta$

$Q_r$ sum of drag torques of rotors in frame $\Gamma_b$

$\tau_\beta$ tilting pitch moment in frame $\Gamma_b$

$\tau_\delta$ moment generated by lift of fixed wings in frame $\Gamma_b$

$\tau_f$ moment generated by free wings in frame $\Gamma_b$

$\tau_{fd}$ moment generated by drag of fixed wings in frame $\Gamma_b$

$\tau_{xd}$ moment generated by drag of free wings in frame $\Gamma_b$

$J_1$ moment of inertia with respect to axis $E_x^\beta$,

$J_2$ moment of inertia with respect to axis $E_y^\beta$

$J_3$ moment of inertia with respect to axis $E_z^\beta$

$J_4$ tilting moment of inertia

**REFERENCES**

[1] T. Oktay, C. Sultan, Modeling and control of a helicopter slung-load system, Aerospace Science and Technology, Vol. 29, No. 1, 2013, 206-222.

[2] C.C. Luo, R.F. Liu, C.D. Yang, Y.H. Chang, Helicopter H∞ control design with robust flying quality, Aerospace Science and Technology, Vol.7, No. 2, 2003, 159-169.

[3] X. Wang, J. Liu, K. Cai, Tracking control for a velocity-sensorless VTOL aircraft with delayed outputs, Automatica, vol. 45, no. 12, 2009, 2876-2882.

[4] X. Wang, Takeoff/landing control based on acceleration measurements for VTOL aircraft, Journal of the Franklin Institute, vol. 350, no. 10, 2013, 3045-3063.

[5] H.L. Tinger, Analysis and application of aircraft departure prediction criteria to the AV-8B Harrier II, AIAA Atmospheric Flight Mechanics Conference, Monterey, CA, Technical Papers; 17-19 Aug. 1987. pp. 343-352.

[6] H. Powrie and A. Novis, Gas path debris monitoring for F-35 Joint Strike Fighter propulsion system PHM, 2006 IEEE Aerospace Conference, Big Sky, MT, July, 2006.

[7] J.G. Leishman, Principles of Helicopter Aerodynamics, Second Edition. Cambridge University Press, 2006.

[8] L.R. Meakin, Moving body overset grid methods for complete aircraft tiltrotor simulations, AIAA Computational Fluid Dynamics Conference, 11th, Orlando, FL, Technical Papers. Pt. 2; 6-9 July 1993. pp. 576-588.

[9] D. Wyatt, Eagle eye pocket guide, Bell Helicopter Textron Inc, Printed in USA, June 2004.

[10] S. M. Barkai, O. Rand, R. J. Peyran, R. M. Carlson, Modelling and analysis of tilt-rotor aeromechanical phenomena, in American Helicopter Society Aeromechanics Specialists Conference, Bridgeport, October 1995.

[11] J. Escareno, S. Salazar, R. Lozano, Modelling and Control of a Convertible VTOL Aircraft, Proceedings of the 45th IEEE Conference on Decision & Control Manchester Grand Hyatt Hotel San Diego, CA, USA, December 13-15, 2006.

[12] R.H. Stone, Control Architecture for a Tail-Sitter Unmanned Air Vehicle, 5th Asian Control Conference, 736-744, 2004.

[13] R.H. Stone, The T-wing tail-sitter unmanned air vehicle: from design concept to research flight vehicle, in Proceedings of the I MECH E Part G Journal of Aerospace Engineering, 2004, pp. 417-433.

[14] N.B. Knoebel, S. R. Osborne, D. Snyder, T. W. McLain, R. W. Beard, A. M. Eldredge, Preliminary Modeling, Control, and Trajectory Design for Miniature Autonomous Tailsitters, AIAA Conference on Guidance,Navigation, and Control, Keystone CO, 2006, paper no. AIAA-2006-6713.

[15] Aviation Week & Space Technology "Aerospace SourceBook 2006,"

[16] AERL Accomplishment Report, May 2004.

[17] Office of the Secretary of Defense, Unmanned aerial vehicles roadmap 2002-2007, December, 2002.

[18] D.J. Taylor, M.V. Ol, T. Cord, Skytote advanced cargo delivery system, in AIAA/ICAS International Air and Space Symposium and Exposition: The Next 100 Years, 2003.

[19] X. Wang, H. Lin, Design and control for rotor-fixed wing hybrid aircraft, Proceedings of the IMechE Part G: Journal of Aerospace Engineering, 2011, vol. 225, no. 7, 831-847.

[20] E. Cetinsoy, S. Dikyar, C. Hancer, K.T. Oner, E. Sirimoglu, M. Unel, and M.F. Aksit, Design and construction of a novel quad tilt-wing UAV, Mechatronics 22 (201) 723-745

[21] C. Hintz, C. Torno, L.R. Garcia Carrillo, Design and dynamic modeling of a rotary wing aircraft with morphing capabilities, 2014 International Conference on Unmanned Aircraft Systems (ICUAS), vol., no., pp.492-498, 27-30 May 2014

[22] C.C.Wang, H.T.Yau, Nonlinear dynamic analysis and sliding mode control for a gyroscope system, Nonlinear Dynamics. 66(1-2), 2011, 53-65.

[23] X. Wang, B. Shirinzadeh, M.H. Ang. Nonlinear double-integral observer and application to quadrotor aircraft, IEEE Transactions on Industrial Electronics, vol. 62, no. 2, 2015, 1189-1200.

[24] X. Wang and B. Shirinzadeh, Nonlinear multiple integrator and application to aircraft navigation, IEEE Transactions on Aerospace and Electronic Systems, vol. 50, no. 1, 2014, 607-622.

[25] X. Wang and B. Shirinzadeh, High-order nonlinear differentiator and application to aircraft control, Mechanical Systems and Signal Processing, vol. 46, no. 2, 2014, 227-252.

[26] M. Knight and R. A. Hefner, Static thrust of the lifting airscrew, NACA TN 626.

[27] R.W. Prouty, Helicopter performance, stability and control, New York: Krieger, 1995.

[28] W.Z. Stepniewsky, Rotor-wing aerodynamics, Basic Theories of Rotor Aerodynamics, New York: Dover, 1984.

[29] W. Johnson, Helicopter Theory, Princeton Univ. Press, Princeton, NJ, pp. 282, 283, 1980.

[30] W. Johnson, Helicopter Optimal Descent and Landing After Power Loss, NASA TM 73244,May 1977.

[31] G. R. Gress, A dual-fan vtol aircraft using opposed lateral tilting for pitch control, in American Helicopter Society 59th Annual Forum, Phoenix, Arizona, May 2003.

[32] S. P. Bhat and D. S. Bemstein, Finite-time stability of continuous autonomous systems, Siam J. Control Optim., 2000, Vol. 38, No. 3, pp. 751-766.

[33] X. Wang, H. Lin, Design and frequency analysis of continuous finite-time-convergent differentiator, Aerospace Science and Technology, vol. 18, no. 1, 2012, 69-78.

[34] A. Levant, High-order sliding modes, differentiation and output-feedback control, International Journal of Control, 2003, vol. 76, Nos. 9/10, 924-941.

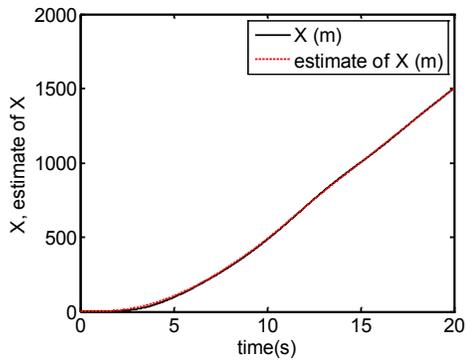
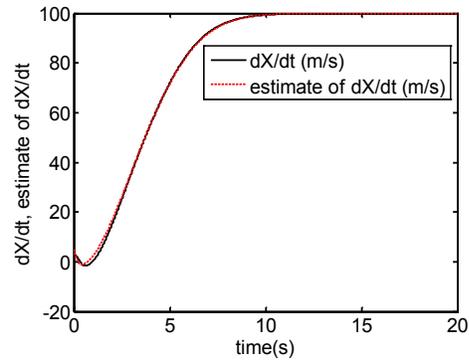

a) Position in X direction

b) velocity in X direction

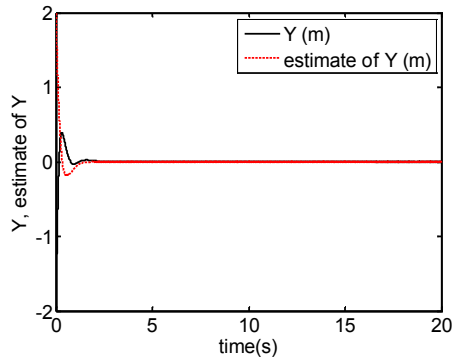
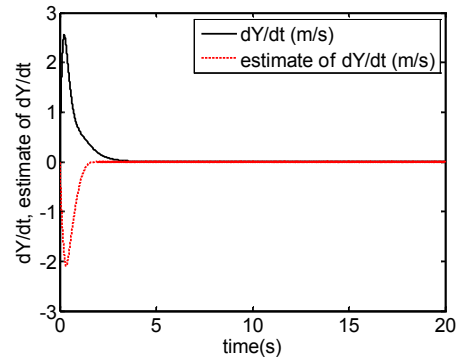

c) Position in Y direction

d) velocity in Y direction

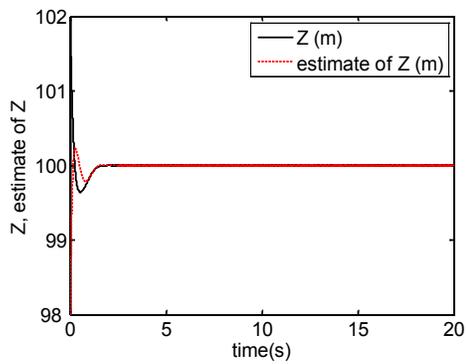
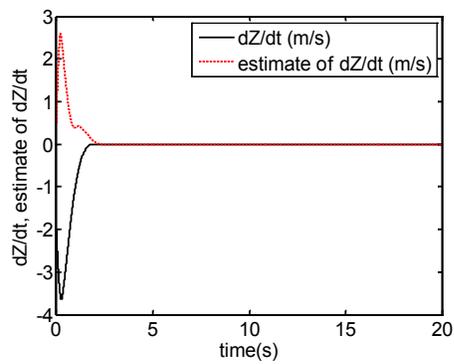

e) Position in Z direction

f) velocity in Z direction

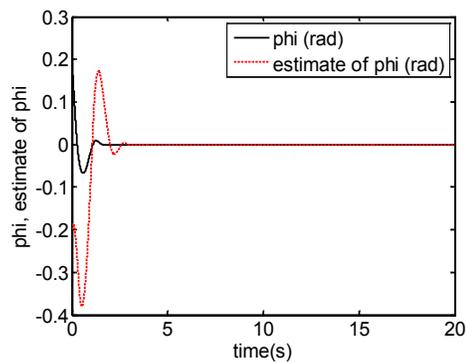
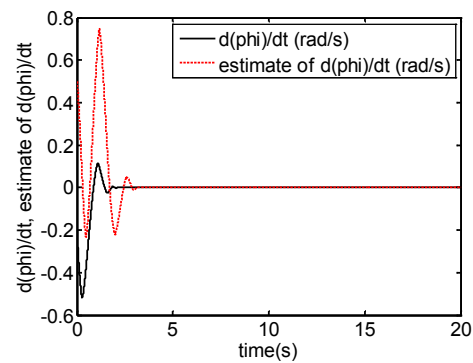

g) Roll angle

h) Roll rate

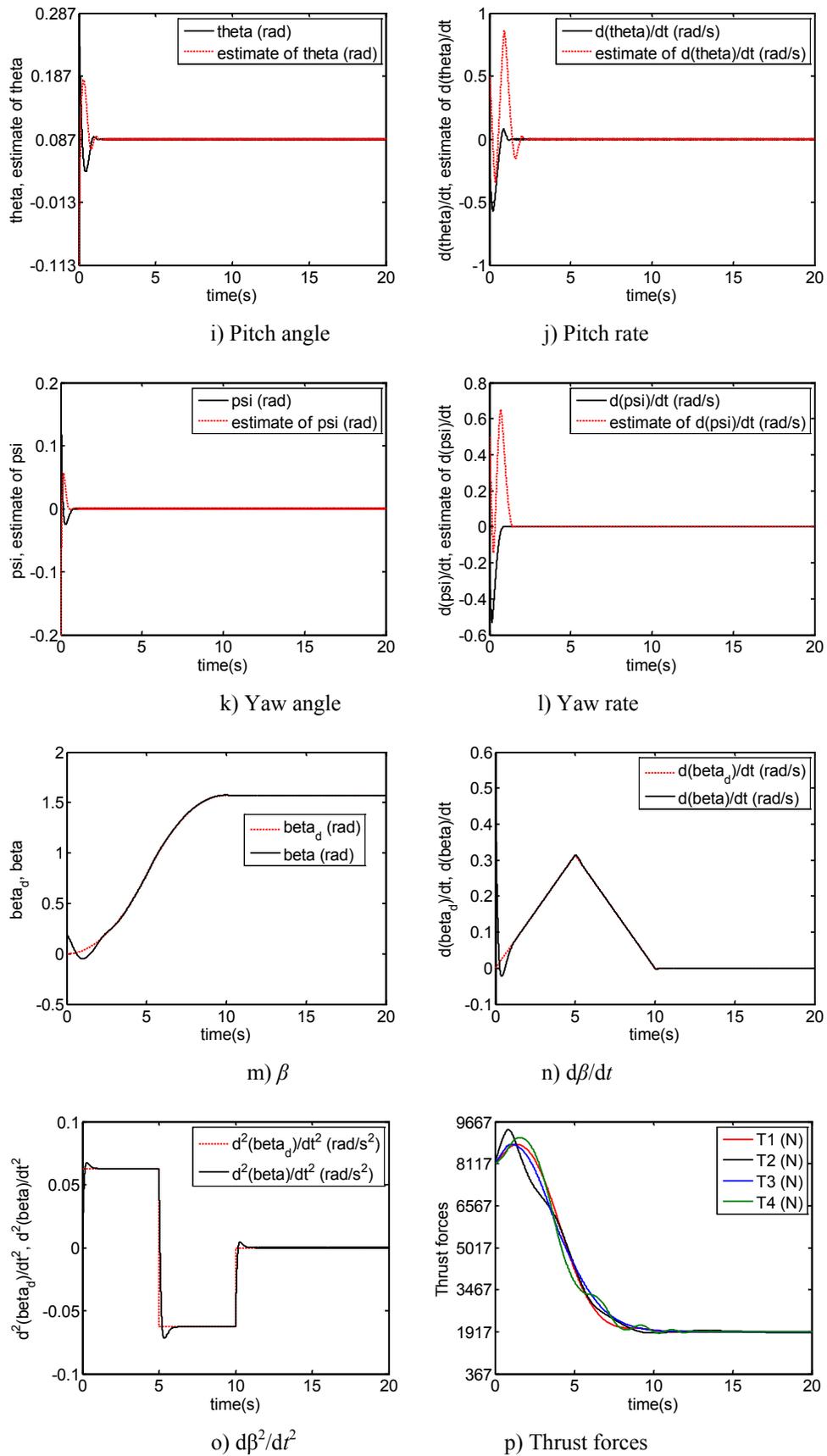

Fig. 7. Hover-to-level transition

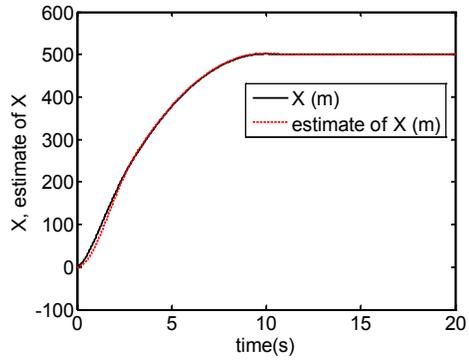

a) Position in X direction

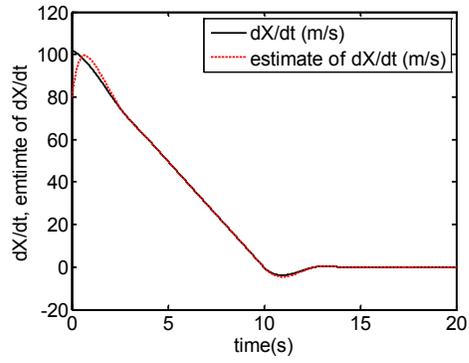

b) velocity in X direction

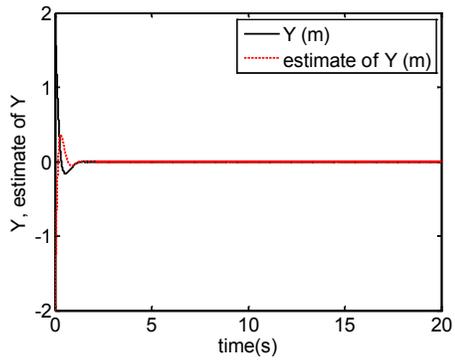

c) Position in Y direction

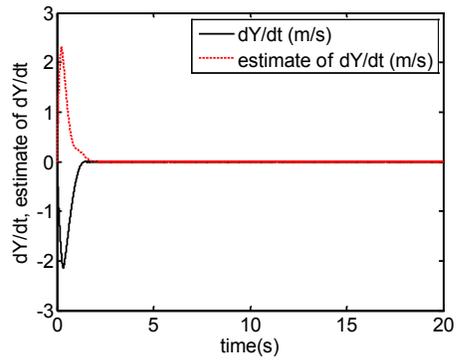

d) velocity in Y direction

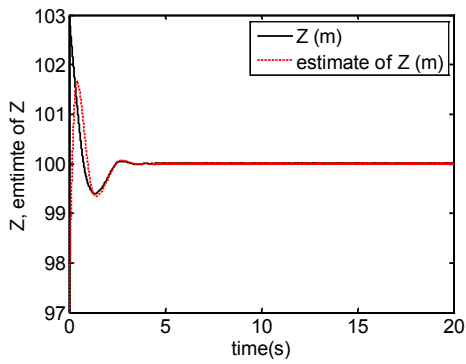

e) Position in Z direction

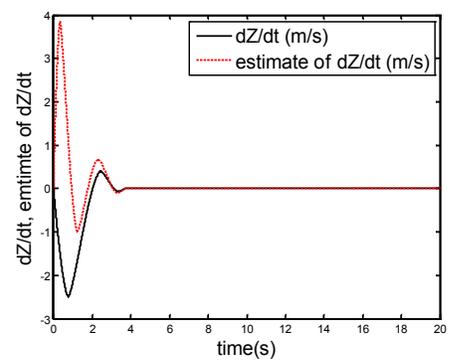

f) velocity in Z direction

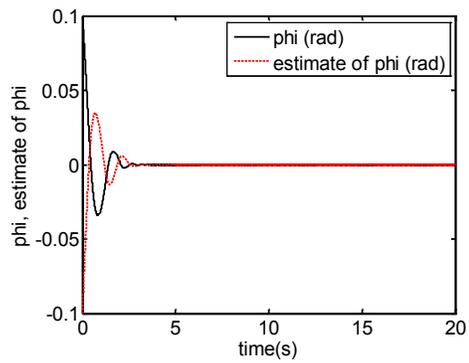

g) Roll angle

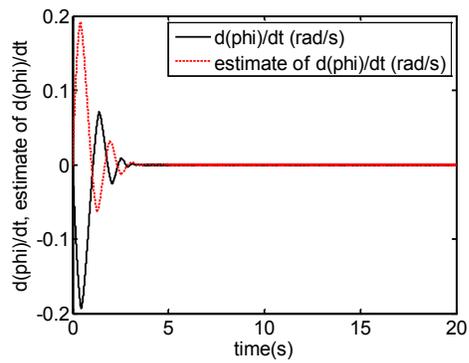

h) Roll rate

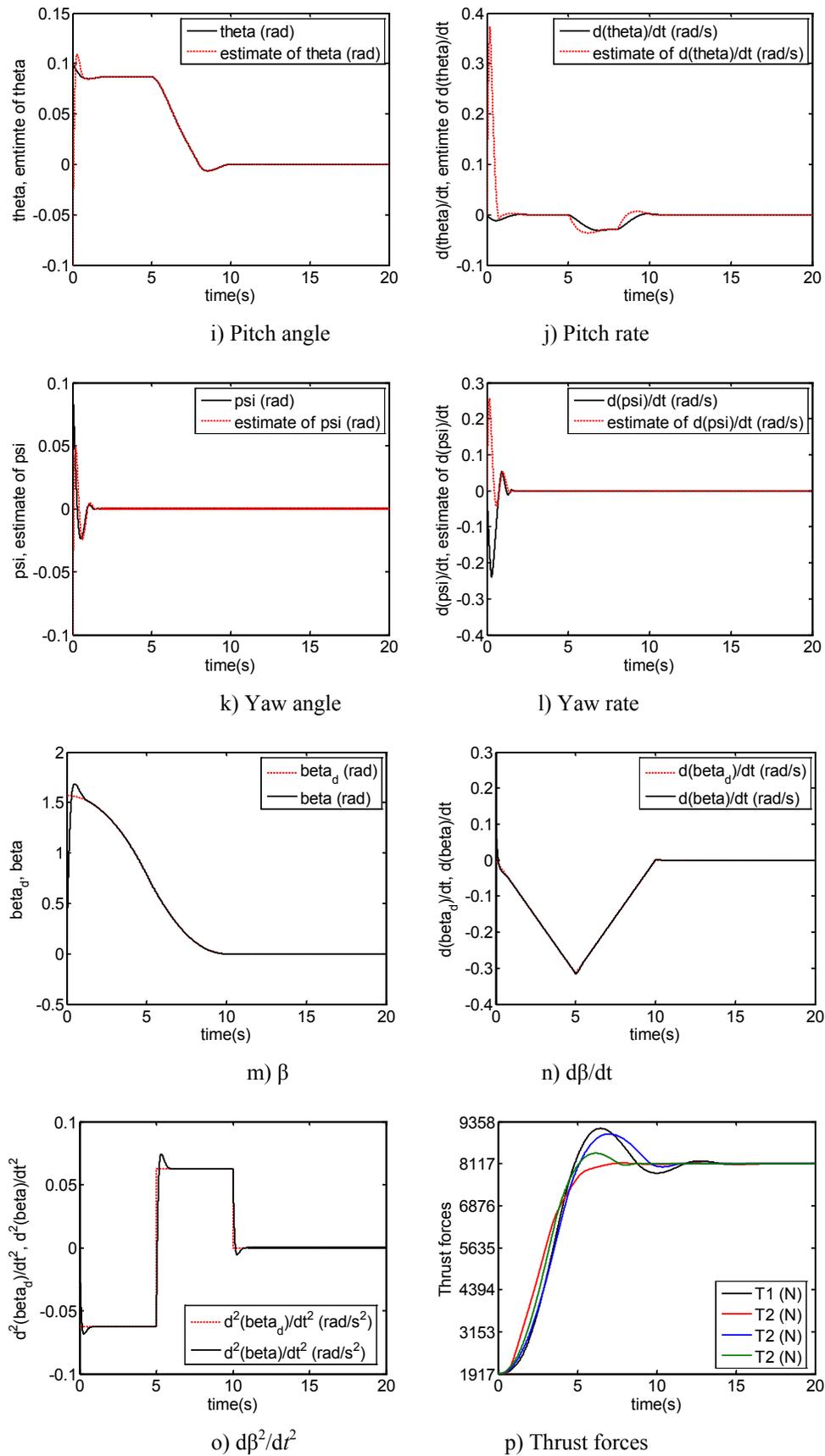

i) Pitch angle

j) Pitch rate

k) Yaw angle

l) Yaw rate

m) β

n) dβ/dt

o) d²β/d$t^2$

p) Thrust forces

Fig. 8. Level-to-hover transition